\documentclass[12pt]{article}
\usepackage{amsmath, amsthm, amsfonts, bm}
\usepackage{graphicx,psfrag,epsf, float}
\usepackage{enumerate,titlesec, color}
\usepackage{natbib}
\usepackage{url} 
\usepackage{soul}

\newcommand{\blind}{0}

\renewcommand\footnotemark{}

\newtheorem{thm}{Theorem}[section]
\newtheorem{theorem}[thm]{Theorem}

\numberwithin{equation}{section}
\theoremstyle{plain}

\theoremstyle{definition}

\usepackage{multirow}
\definecolor{napiergreen}{rgb}{0.16, 0.5, 0.0}
\definecolor{myrtle}{rgb}{0.13, 0.26, 0.12}
\definecolor{dartmouthgreen}{rgb}{0.05, 0.5, 0.06}

\newcommand{\sumi}{\sum_{i=1}^n}
\newcommand{\mZ}{\mathcal{Z}}

\newcommand{\mN}{\mathcal{N}}
\newcommand{\mO}{\mathcal{O}}

\newcommand{\mU}{{\mathcal U}}
\newcommand{\bbeta}{{\bm\beta}}
\newcommand{\balpha}{{\bm\alpha}}
\newcommand{\bgamma}{{\bm\gamma}}
\newcommand{\btheta}{{\bm\theta}}
\newcommand{\bU}{{\bm U}}
\newcommand{\bS}{{\bm S}}
\newcommand{\bB}{{\bm B}}
\newcommand{\bet}{{\bm \eta}}
\newcommand{\bD}{{\bm D}}

\newcommand{\bphi}{{\bm \phi}}
\newcommand{\bs}{{\bm s}}
\newcommand{\bE}{{\bm {\mathcal E}}}
\newcommand{\bZ}{{\bm Z}}
\newcommand{\bSigma}{{\bm \Sigma}}
\newcommand{\bX}{{\bm X}}
\newcommand{\bW}{{\bm W}}
\newcommand{\ba}{{\bm a}}
\newcommand{\bpsi}{{\bm \psi}}
\newcommand{\bzero}{{\bm 0}}
\begin{document}

	\def\spacingset#1{\renewcommand{\baselinestretch}%
	{#1}\small\normalsize} \spacingset{1}

\if0\blind
{	\title{Recurrent Events Analysis With Data Collected at Informative Clinical Visits in Electronic Health Records}
	\date{}
	\author{Yifei Sun, Charles E.\ McCulloch, Kieren A.\ Marr, and Chiung-Yu Huang
	\footnote{Yifei Sun is Assistant Professor, Department of Biostatistics, Columbia University Mailman School of Public Health, New York, NY 10032 (Email: \emph{ys3072@cumc.columbia.edu}).  Charles E.\ McCulloch is Professor, Department of Epidemiology and Biostatistics, University of California San Francisco, San Francisco, CA 94158 (Email: \emph{Charles.McCulloch@ucsf.edu}). Kieren A.\ Marr is Professor, Johns Hopkins University School of Medicine, Baltimore, MD 21205 (Email: \emph{kmarr4@jhmi.edu}). Chiung-Yu Huang is Professor, Department of Epidemiology and Biostatistics, University of California San Francisco, San Francisco, CA 94158 (Email: \emph{ChiungYu.Huang@ucsf.edu}).}}
	\maketitle
} \fi

\if1\blind
{
	\bigskip
	\date{}
	\title{Recurrent Events Analysis With Data Collected at Informative Clinical Visits in Electronic Health Records}
	\maketitle
} \fi

\def\spacingset#1{\renewcommand{\baselinestretch}%
	{#1}\small\normalsize} \spacingset{1}
\spacingset{1.45}
\vspace{-0.18in}
\begin{abstract} 
Although increasingly used as a data resource for assembling cohorts, electronic health records (EHRs) pose many analytic challenges. In particular, a patient's health status influences when and what data are recorded, generating sampling bias in the collected data. In this paper, we consider recurrent event analysis using EHR data. Conventional regression methods for event risk analysis usually require the values of covariates to be observed throughout the follow-up period. In EHR databases, time-dependent covariates are intermittently measured during clinical visits, and the timing of these visits is informative in the sense that it depends on the disease course. Simple methods, such as the last-observation-carried-forward approach, can lead to biased estimation. On the other hand, complex joint models require additional assumptions on the covariate process and cannot be easily extended to handle multiple longitudinal predictors. By incorporating sampling weights derived from estimating the observation time process, we develop a novel estimation procedure based on inverse-rate-weighting and kernel-smoothing for the semiparametric proportional rate model of recurrent events. The proposed methods do not require model specifications for the covariate processes and can easily handle multiple time-dependent covariates. Our methods are applied to a kidney transplant study for illustration. 
\end{abstract}
\noindent%
\vfill

\section{Introduction}

The electronic health record (EHR) is a longitudinal, digital record of a patient's health-related information generated at medical encounters in any healthcare setting. The EHR contains clinical data that may be absent from insurance claims, such as vital signs, laboratory results, and imaging reports, and thus can provide new insights into risk factors and natural history of a disease. In recent years, EHR databases have been increasingly utilized as a resource for assembling study cohorts for clinical research as they provide a low-cost means of accessing rich longitudinal data in large samples \citep{fiks2012comparative,coorevits2013electronic,van2016feasibility}.  Unlike clinical trials and epidemiological cohort studies where data collection is driven by research needs, EHR data are collected for clinical care and billing purposes. In an EHR, a patient's health status influences the frequency of medical encounters, and the disease course can affect the type of clinical assessments used, lab tests ordered, and imaging performed at each visit. Such a data collection scheme leads to sparse, irregular sampling, that is, the time between clinical visits varies within a patient and across patients \citep{goldstein2016controlling,casey2016using}. Moreover, the timing of the visits is potentially informative and associated with the outcome of interest, posing analytical challenges for conducting epidemiologic research using EHR-based cohorts \citep{luo2013methods,phelan2017illustrating}.

This research was motivated by an IRB-approved observational study of recurrent infections in a cohort of 160 patients who underwent a kidney transplant at the Johns Hopkins Hospital in the year of 2012. As serious infections are known to contribute to organ loss and morbidity after transplant \citep{humar2010efficacy, razonable2013cytomegalovirus}, understanding infection risk allows for the development and implementation of better informed, personalized prevention strategies. The data include all the post-transplant visits extracted from medical records. Our analysis focuses on serious bacterial, viral and fungal infections at post-transplant visits that were identified by applying objective laboratory and clinical criteria. The clinical visits at which serious infections were determined are termed \emph{event visits}, while the others are termed \emph{non-event visits}.  In addition to baseline demographic and clinical factors, we are interested in the effect of serum creatinine level, as a time-dependent covariate, on the infection risk. In this study, serum creatinine level was measured at both event and non-event visits to monitor the kidney function after transplant. Patients with human leukocyte antigen (HLA) incompatible or deceased donor were monitored more closely, because they are at a higher risk of graft failure. These irregular and {subject-specific} visit times create substantial challenges for valid statistical inference.

To estimate the effect of risk factors on serious infections, we postulate a semiparametric proportional rate model \citep{lin2000semiparametric} for the counting process of recurrent infections. This model is essentially an extension of Cox's proportional hazards model to the setting of recurrent events and, analogously, it allows the baseline rate function to be unspecified. Fitting the proportional rate model requires the covariate processes to be continuously monitored. However, in EHR data, time-dependent covariates are only observed at clinical visits. A common strategy is to impute the unobserved covariate values by the last-observation-carried-forward (LOCF) approach. Under LOCF, the last known value of the covariate is used forward in time until a new value is measured. Thus the true covariate process is approximated by a step function with jumps at the observation times. Despite its simplicity, the LOCF method is known to yield biased estimation in survival and recurrent event analysis \citep{prentice1982covariate, faucett1998analysis, cao2015analysis, li2016recurrent}.

As an alternative to LOCF, joint modeling approaches which simultaneously model the event processes and the longitudinal measurements have been extensively studied in the literature; see \cite{tsiatis2004joint} and \cite{rizopoulos2012joint} for comprehensive overviews. In the presence of irregular and  informative observation times, researchers, including \cite{liu2008analysis}, \cite{sun2012joint}, \cite{han2014joint}, \cite{li2016joint}, \cite{dai2017joint}, have considered joint models of survival and longitudinal data, where the sub-models are linked via latent variables. The latent variable approach, although well-developed, does not give a direct interpretation of the effects of time-dependent covariates (i.e., the longitudinal variable) on event risk, because the event time model usually involves the latent variables rather than the intermittently observed, time-dependent covariates. Moreover, the existing joint models usually deal with a single longitudinal variable, whose distribution is often assumed to be continuous. Extensions to handling multiple, mixed-type time-dependent covariates are not straightforward. Finally, the validity of the inference on event risk relies heavily on correct specification of the longitudinal model \citep{yao2007functional,li2016recurrent}, which is especially challenging when dealing with a large number of longitudinal variables. To the best of our knowledge, no existing joint modeling approach provides a satisfactory solution to handle multiple and mixed type time-dependent covariates that can potentially arise in EHR research.

To deal with intermittently observed, time-dependent covariates in Cox-type models, there has been a rising interest in developing estimation procedures based on estimated partial likelihood score functions via kernel smoothing. For example, \cite{cao2015analysis} considered the case where the time-dependent covariates are measured at noninformative, irregular follow-up visits but not at event times, and proposed a kernel weighted score function approach. \cite{li2016recurrent} focused on the case where the time-dependent covariates are observed at both event occurrences and noninformative irregular follow-up visits. Compared to the joint modeling approaches, the aforementioned methods do not postulate a model for the longitudinal covariates and can easily handle multiple covariates (continuous and/or categorical). However, the consistency of the kernel-smoothing methods relies on a key assumption that the non-event follow-up visits are noninformative, which is often violated in the analysis of EHR data.

In this article, we develop novel semiparametric methods to estimate the proportional rate model with intermittently observed time-dependent covariates from EHR data. The proposed approach does not require a model for the covariate processes and can deal with multiple time-dependent covariates. Compared to the existing kernel smoothing approaches, our method allows the non-event visits to depend on the recurrent event outcome as well as the time-dependent covariates through the observed history. Specifically, by exploring the mechanisms and assumptions of visit time processes, we propose to model the non-event visit time process and estimate the unknown complete-data partial likelihood score function using inverse-rate-weighting and kernel smoothing.

The article is organized as follows. In Section 2, we introduce the main model on the recurrent event outcome; we further show how the model assumptions are integrated to provide identification of the parameters of interest. In Section 3, we propose a two-step estimation procedure based on inverse-rate-weighting, and investigate the proposed method when the observation times are informative. In Section 4, simulation studies are carried out to evaluate the performance of the proposed method. In Section 5, the proposed methods are applied to the kidney transplant data. We conclude the paper with a discussion in Section 6.

\section{Model Setup}

{In our framework, we distinguish visits when an event has been declared to have occurred from non-event visits. In our example, an event visit occurs when a serious infection has been detected by applying laboratory and clinical criteria. Our focus for inference is the event visit process. In Section 2.1, we begin by reviewing the popular proportional rate model for the recurrent event outcome (i.e., the event visits). Although time-dependent covariates are measured at all visits, in Section 2.2, we propose to approximate the covariate mean function in the partial score equation of the event model using covariates observed at the non-event visits but not the event visits, as the inclusion of event visits is likely to introduce estimation bias. We discuss different mechanisms of non-event visits and specify a Cox-type  model, under which the regression parameters in the main recurrent event model can be consistently estimated.}

\subsection{The recurrent event outcome model}
Let $N^*(t)$ denote the number of event visits occurring in the time interval $(0,t]$, and $dN^*(t)$ gives the number of events in $[t,t+dt)$. Let $\bZ(t)$ be a $p$-dimensional vector of possibly time-dependent covariates, and $\bZ(t)$ may include baseline covariates whose paths are constant over time. We assume that $\bZ(t)$ is left-continuous and has a right-hand limit. The goal is to estimate the rate function of $N^*(t)$ given $\bZ(t)$, that is, 
$$\mu\{t\mid \bZ(t)\}dt = E\{dN^*(t)\mid \bZ(t) \}.$$ 
The rate function $\mu\{t\mid \bZ(t)\}$ characterizes the instantaneous risk of event occurrence without conditioning on the preceding event history. This is in contrast with the intensity function, which is defined as the instantaneous risk of event occurrence conditioning on the event history. The intensity function uniquely determines the probability structure of the event process, while the rate function allows for arbitrary dependence structure among the recurrent events. When the research interest is to identify risk factors or evaluate treatment effects, modeling the rate function is preferred because the regression parameters have a direct marginal interpretation on the recurrent event risk \citep{cook2007statistical}.
 
We postulate the proportional rate model  \citep{lin2000semiparametric} for the recurrent event outcomes,
\begin{align}
\label{p2}
\mu\{t\mid \bZ(t)\} =\mu_0(t) \exp\{\bbeta_0^\top \bZ(t)\},
\end{align} 
where $\mu_0(t)$ is an unspecified baseline rate function and $\bbeta_0$ is a $p\times 1$ vector of regression coefficients. The $j$th component of the regression parameter gives the log ratio of the rate function at time $t$ for every unit increase in the $j$th covariate. The observation of recurrent events is usually subject to right-censoring due to loss to follow-up or study end. We denote the censoring time by $C$. The observed event counting process is $N(t) = N^*(t \wedge C)$, where $a \wedge b = \min(a,b)$. In the ideal case where we have continuous monitoring of the covariate process $\bZ(t)$ during the follow-up period, the observed data are $\{(N_i(t),\bZ_i(t)),t\le C_i, i=1,\ldots,n \}$. Following \cite{lin2000semiparametric}, the regression coefficient $\bbeta_0$ can be consistently estimated by solving the pseudo-partial score equation $\bU(\bbeta)=\bzero$, where
\begin{align}
\bU(\bbeta) 
=\frac{1}{n}\sumi \int_{0}^{\tau} \bZ_i(t) dN_i(t) -\frac{1}{n}\sumi \int_{0}^{\tau} \frac{\bS^{(1)}(t,\bbeta)}{S^{(0)}(t,\bbeta)} dN_i(t),  
\label{ee_partial}
\end{align}
$\bS^{(k)}(t,\bbeta) = n^{-1}\sumi I(C_i\ge t)\bZ_i(t)^{k} \exp\{\bbeta^\top \bZ_i(t) \}$, and $\tau$ is a predetermined constant {such that the event process potentially could be observed up to $\tau$}. The evaluation of  ${\bS^{(1)}(t,\bbeta)}/{S^{(0)}(t,\bbeta)}$ requires $\bZ(t)$ to be known for all the subjects who are under follow-up at $t$ (i.e., $C\ge t$).

As pointed out in \cite{li2016recurrent}, the score function $\bU(\bbeta)$ can be viewed as a functional of the four stochastic processes $n^{-1}\sumi dN_i(t)$, $n^{-1}\sumi \bZ_i(t) dN_i(t)$, $S^{(0)}(t,\bbeta)$, and $\bS^{(1)}(t,\bbeta)$. Denote by $\bs^{(k)}(t,\bbeta) = E[I(C\ge t)\bZ(t)^{k} \exp\{\bbeta^\top \bZ(t)\} ]$ the limiting function of $\bS^{(k)}(t,\bbeta)$, $k=0,1$.   
Since the mapping defined by $\bU(\bbeta)$ is compactly differentiable,  $\bU(\bbeta)$ converges in probability as $n\rightarrow \infty$ to the corresponding functional, that is,
$$\bU(\bbeta)\overset{p}{\longrightarrow} {\bm \mU}_0(\bbeta) =  \int_{0}^{\tau}E\left\{\bZ(t)dN(t)\right\}-\int_{0}^{\tau}\bE(t,\bbeta)E\{dN(t)\},$$ 
where $\bE(t,\bbeta) = {\bs^{(1)}(t,\bbeta)}/{s^{(0)}(t,\bbeta)}$. Under the proportional rate model (\ref{p2}), one has ${\bm \mU}_0(\bbeta_0)=\bzero$. Heuristically, if another estimator that converges in probability to $\bE(t,\bbeta)$ could be substituted for $\bS^{(1)}(t,\bbeta)/S^{(0)}(t,\bbeta)$ in (\ref{ee_partial}), solving the modified estimating equation will yield a consistent estimator of $\bbeta_0$.

As shown below, one can estimate $\bE(t,\bbeta)$ using $\bZ(\cdot)$ observed from the non-event visits. Let $O^*(t)$ be the counting process for non-event visits, and denote by $O(t) =O^*(t\wedge C)$ the counting process for non-event visits during the follow-up period. \cite{li2016recurrent} proposed the kernel smoothing estimator
\begin{align}
\label{k1}
\widehat{\bE}_O({t},\bbeta) = \frac{\int_{0}^{\infty}K_h(t-s) \widehat{E}[\bZ(s)\exp\{\bbeta^\top \bZ(s)\} dO(s)]}{\int_{0}^{\infty}K_h(t-s) \widehat{E}[\exp\{\bbeta^\top \bZ(s)\} dO(s)]},
\end{align}
where $\widehat{E}[\bZ(s)^{k} \exp\{\bbeta^\top \bZ(s)\} dO(s)] = n^{-1} \sumi \bZ_i(s)^{k} \exp\{\bbeta^\top \bZ_i(s)\} dO_i(s)$ are the corresponding empirical estimates for $k = 0,1$, $K(\cdot)$ is a symmetric kernel function with support $[-1,1]$, $h$ is a bandwidth parameter, and $K_h(\cdot) = K(\cdot/h)/h$. Under the assumption that $O^*(\cdot)$ is independent of $\{\bZ(\cdot),C \}$,  one can show that $\widehat{\bE}_O(t,\bbeta)$ converges in probability to $\bE(t,\bbeta)$ as $n\rightarrow\infty$. As a result, solving $n^{-1}\sumi \int_{0}^{\tau} \{ \bZ_i(t)-\widehat{\bE}_O(t,\bbeta) \}dN_i(t)=\bzero$ is expected to yield a consistent estimator of $\bbeta_0$.

The independence assumption on the timing of non-event visits, however, is usually violated in EHR data. In this case, the kernel smoothing method can yield biased estimation. In Section 2.2, we introduce a {Cox-type} regression model for the non-event visits and weight the kernel smoothing estimator accordingly to obtain a consistent estimator of $\bE(t,\bbeta)$.

\subsection{The non-event visit time model}

In what follows, we investigate ways to use the non-event visits to facilitate the estimation of the covariate mean function $\bE(t,\beta)$. In EHR data, the timing of a non-event visit is often correlated with the covariate processes and the event visit process. To model the non-event visits, one needs to consider the interplay between the two counting processes $O^*(\cdot)$ and $N^*(\cdot)$ and their dependency on $\bZ(\cdot)$. We begin by considering a simple case where the decision to schedule a non-event visit is made independently of the time-dependent covariates. We term this assumption as visiting completely at random (VCAR). Specifically, VCAR assumes that 
$$E\{dO^*(t)\mid {\bZ}(t) \} = \lambda(t)dt,$$ where $\lambda(t)$ is a function that does not depend on $\bZ(t)$. Note that the method in \cite{li2016recurrent} can be applied to consistently estimate $\bbeta_0$ under VCAR.

In reality, the chance of having a non-event visit can be related to time-varying risk factors as well as the recurrent event process. In EHR research, the timing of the next non-event visit often depends on the clinical assessments or lab results at the previous visits.  For example, physicians may request a check-up visit in the near future if an elevated biomarker level was detected at the previous visit. In what follows,  we relax the VCAR assumption to allow for dependency between $O^*(\cdot)$ and $\bZ(\cdot) $ through observed history information. Denote by $\mO^*(t) = \{O^*(u),0\le u<t \}$, $\mN^*(t) = \{N^*(u),0\le u<t \}$ and ${\bm \mZ}^*_{\rm obs}(t) = \{(\bZ(u)dN^*(u), \bZ(u)dO^*(u)),  0\le u< t\}$ the observed history of event visits, non-event visits and covariates prior to $t$. Let $\bX(t)$ be a $q$-dimensional vector of the observed medical history prior to $t$, that is, $\bX(t)$ is defined as $\bX(t) = \phi(\mO^*(t), \mN^*(t), {\bm \mZ}^*_{\rm obs}(t))$, with $\phi$ being a pre-specified function. For example, $\bX(t)$ may include the observed value of $\bZ(t)$ from the last visit before time $t$ as well as time-independent patient characteristics that can affect the visiting frequency. In the spirit of the missing at random (MAR) assumption in the missing data literature \citep{tsiatis2007semiparametric}, we give the following definition of {visiting at random} (VAR), 
\begin{align}
\label{a1}
E\{d{O}^*(t) \mid {\bZ}(t), \bX(t)\} = E\{d{O}^*(t) \mid \bX(t)\}.
\end{align}
Assumption  (\ref{a1}) implies that the decision to schedule a non-event visit at time $t$ depends on $\{\bZ(t),\bX(t) \}$ only through the values of $\bX(t)$ that are available to the investigators.

The VAR assumption allows us to construct an inverse-rate-weighted estimator for the limiting function $\bE(t,\bbeta)$. We postulate a Cox-type model for the non-event visit counting process,
\begin{align}
\label{m2}
E\{dO^*(t)\mid\bX(t)\}  =\exp\{\balpha_0^\top \bX(t) \}\lambda_0(t)dt,
\end{align}
where $\lambda_0(t)$ is an unspecified baseline function, and $\balpha_0$ is a $q\times 1$ vector of regression parameters. We next consider estimating $\bE(t,\bbeta)$ under (\ref{a1}) and (\ref{m2}). In the presence of censoring, we further assume independent censoring in the sense that
$E\{d{O}^*(t) \mid \bZ(t),\bX(t),  C\ge t\}=E\{d{O}^*(t) \mid \bZ(t), \bX(t)\}.$
Our estimation is motivated by the following result: for $k = 0,1$, by the law of total expectation, we have
\begin{eqnarray}
\label{key}
\lefteqn{ E\left[\bZ(t)^{k} \exp\{\bbeta^\top \bZ(t) - \balpha_0^\top \bX(t)\} d{O}(t) \right] }\nonumber\\
&=& E\left[I(C\ge t)\bZ(t)^{k} \exp\left\{\bbeta^\top \bZ(t) - \balpha_0^\top \bX(t) \right\} E\left\{d{O}^*(t)\mid \bZ(t),{\bX}(t),  C\ge t \right\}\right]\nonumber\\
&=& E\left[ I(C\ge t)\bZ(t)^{k} \exp\left\{\bbeta^\top \bZ(t) - \balpha_0^\top \bX(t) \right\}  \exp\left\{ \balpha_0^\top \bX(t) \right\}\lambda_0(t)dt\right]\nonumber\\
&=& \bs^{(k)}(t,\bbeta)\lambda_0(t)dt.
\end{eqnarray}
It follows from Equation $(\ref{key})$ that, if the value of $\balpha_0$ is known, the limiting function $\bE(t,\bbeta)$ can also be estimated via kernel smoothing by replacing $dO(s)$ with $ \exp\{-\balpha_0^\top \bX(s) \}dO(s)$ in (\ref{k1}). Thus $\bbeta_0$ can be consistently estimated by solving the modified score equation, with $\bS^{(1)}(t,\bbeta)/S^{(0)}(t,\bbeta)$ replaced with the new kernel type estimator. In practice, $\balpha_0$ can be estimated from the data. A detailed description of the estimation procedure is given in Section 3.1.

Finally, if Assumption (\ref{a1}) does not hold, that is, conditioning on the observed history information $\bX(t)$, the current value of $\bZ(t)$ has additional effects on the chance of a non-event visit at $t$, we define this type of visiting mechanism to be \textit{visiting not at random} (VNAR). Borrowing information from covariates observed at this type of visit can lead to biased estimation of $\bE(t,\bbeta)$. In Section 3.2, we investigate the degree of bias of the kernel smoothing estimators and propose potential solutions in several special cases of VNAR. We note that \cite{pullenayegum2016longitudinal} used the same terminology in the classification of visit processes in the context of longitudinal data analysis, but our definitions are different since we take both the event and non-event visits into account.

\section{Model Estimation}
\subsection{Estimation under VAR}
In this section, we present the proposed estimation procedure under VAR and Model (\ref{m2}). The observed data $\{ C_i, \mN^*_i(C_i),\mO^*_i(C_i), {\bm \mZ}^*_{{\rm obs},i}(C_i);$ $  i=1,\dots,n\}$ are assumed to be $n$ independent realizations of $\{ C, \mN^*(C),\mO^*(C), {\bm \mZ}^*_{{\rm obs}}(C)\}$. 
As discussed in Section 2.1,  solving (\ref{ee_partial}) for 0 with ${\bS}^{(1)}(t,\bbeta)/{S}^{(0)}(t,\bbeta)$ being replaced with an estimator that converges to the same limiting function would give a consistent estimator. Hence our goal is to estimate the limiting function $\bE(t,\bbeta)$ using data observed at non-event visits. For $k = 0,1$, define 
\begin{align}
\label{est1}
& \widehat{\bS}^{(k)}(t,\bbeta,\balpha) =  \int_0^\infty K_h(t-s) \widehat{E}[\bZ(s)^{k} \exp\{\bbeta^\top \bZ(s) - \balpha^\top \bX(s)\} dO(s)],
\end{align}
where $\widehat{E}$ denotes the empirical estimate. Based on Equation (\ref{key}), when $n\rightarrow\infty$, $h\rightarrow 0$ and $nh\rightarrow\infty$, $\widehat{\bS}^{(k)}(t,\bbeta,\balpha_0)$ consistently estimates $\bs^{(k)}(t,\bbeta)\lambda_0(t)$ for a given $\bbeta$.
Thus if $\balpha_0$ were known, ${\bE(t,\bbeta)}$ can be consistently estimated by
\begin{align*}
& \widehat{\bE}(t,\bbeta, \balpha_0) = \frac{\widehat{\bS}^{(1)}(t,\bbeta,\balpha_0)}{\widehat{S}^{(0)}(t,\bbeta,\balpha_0)}.
\end{align*}
For the $i$th subject, let $V_{ik}$, $k = 1,\ldots,m_i$, be the time to the $k$th non-event visit, {where $m_i$ is the number of observed non-event visits}. Define $w_{ik}(t) = K_h(t-V_{ik})\exp\{-\balpha_0^\top \bX_i(V_{ik}) \}$, then $\widehat{\bE}(t,\bbeta, \balpha_0)$ can be re-expressed as 
$$\frac{\sum_{i=1}^n \sum_{k=1}^{m_i}w_{ik}(t)\exp\{\bbeta^\top \bZ_i(V_{ik}) \}\bZ_i(V_{ik})}{\sum_{i=1}^n \sum_{k=1}^{m_i}w_{ik}(t)\exp\{\bbeta^\top \bZ_i(V_{ik}) \}}.$$ 
Similar to the estimator in (\ref{k1}), the weight $w_{ik}(t)$ depends on the distance between $V_{ik}$ and $t$.  Furthermore, the weight is inversely proportional to the ``risk'' that the covariate value is observed at $V_{ik}$. Under VAR, the chance of being observed near time $u$ can vary across subjects. To obtain a consistent estimate of the limiting function $\bE(t,\bbeta)$, one needs to downweight the observed covariates from subjects who are more likely to visit. Therefore, $ \widehat{\bE}(t,\bbeta, \balpha_0)$ can be viewed as an inverse-rate-weighted kernel type estimator. For $t\in[0,h)$, we set $\widehat{\bE}(t,\bbeta,\balpha_0) = \widehat{\bE}(h,\bbeta,\balpha_0)$ to correct the biased estimation near the boundary.

In practice, the value of $\balpha_0$ is unknown. Under Model (\ref{m2}), we estimate $\balpha_0$ using data from the non-event visits. Specifically, we obtain the estimator for $\balpha_0$ by solving the partial score equation \citep{andersen1982cox,lin2000semiparametric} 
\begin{align}
\label{ee1}
\bU_1(\balpha) =\frac{1}{n} \sum_{i=1}^{n} \int_{0}^{\tau} \{\bX_i(u) - \bar{\bX}(u,\balpha)\} d{O}_i(u) = \bzero,
\end{align}
where $$\bar{\bX}(u,\balpha) = \frac{\sum_{i=1}^{n}I(C_i\ge u) \bX_i(u)\exp\{\balpha^\top \bX_i(u) \}}{\sum_{i=1}^{n}I(C_i\ge u)\exp\{\balpha^\top \bX_i(u) \}}.$$ 
Let $\widehat{\balpha}$ denote the solution of equation (\ref{ee1}). It is shown in the Appendix that $\widehat{\balpha}$ is a consistent estimator for $\balpha_0$, and $\sqrt{n}(\widehat{\balpha}-\balpha_0)$ converges in distribution to a zero mean normal distribution as $n\rightarrow\infty$. In this way,  $\widehat{\bE}(t,\bbeta,\widehat{\balpha})$ naturally serves as an estimator for $\bE(t,\bbeta)$. It is worthwhile to point out that although the estimation procedure uses the inverse-rate-weighting technique, we do not need to explicitly estimate the baseline function $\lambda_0(t)$ when estimating $\bbeta_0$.

Based on the above results, we propose a two-step procedure to estimate $\bbeta_0$. In the first step, we estimate $\balpha_0$ in the non-event visit time model with $\widehat{\balpha}$. In the second step, we replace ${\bS}^{(1)}(t,\bbeta)/{S}^{(0)}(t,\bbeta)$ in (\ref{ee_partial}) with $\widehat{\bE}(t,\bbeta,\widehat{\balpha})$, and solve the following estimating equation, 
\begin{align}
\label{ee2}
\bU_2(\bbeta) = \frac{1}{n}\sum_{i=1}^{n} \int_{0}^{\tau} \{\bZ_i(t) - \widehat{\bE}(t,\bbeta,\widehat{\balpha})\} dN_i(t) = \bzero.
\end{align}
Let $\widehat{\bbeta}$ denote the solution of $\bU_2(\bbeta)=\bzero$. Theorem \ref{th1} summarizes the large sample properties of $\widehat{\bbeta}$. Although the estimation procedure involves kernel smoothing, when the bandwidth $h \propto n^{-\nu}$, $1/4<\nu<1/2$, the estimator achieves the regular $\sqrt{n}$ convergence rate. 
\begin{theorem}
	\label{th1}
	Under regularity conditions (A1)-(A8) in the Appendix, $\sqrt{n}(\widehat{\bbeta}-\bbeta_0)$ converges in distribution to a zero mean normal distribution $N({\bm 0},\bSigma)$, where $\bSigma$ is defined in the Appendix.
\end{theorem}
For the problem of bandwidth selection, the goal is to get a bandwidth $h$ such that $h \propto n^{-\nu}$, $1/4<\nu<1/2$. Note that a precise target of an optimal bandwidth is often unnecessary. For example, if another bandwidth selection procedure is known to yield an optimal bandwidth $h_0 = c_0n^{-\nu_0}$ and $c_0$ is a constant that does not depend on $n$, we can use $h = c_0n^{-1/3}$ as the bandwidth \citep{maity2007efficient}. One may also follow the bandwidth selection procedure described in \cite{li2016recurrent} by choosing $h$ based on the pseudo partial likelihood. In practice, we  recommend trying different bandwidths to evaluate their impact on the $\bbeta$ estimates.

When $\bZ(\cdot)$ is continuously monitored, the baseline cumulative rate function $\mathcal{M}_0(t) = \int_{0}^t \mu_0(u)du$ can be consistently estimated by the following Breslow-type estimator,
\begin{align*}
\widetilde{\mathcal{M}}_0(t) = \frac{1}{n}\int_{0}^t \frac{\sum_{i=1}^n dN_i(u)}{ S^{(0)}(u,\widehat{\bbeta}) }.
\end{align*} 
The process $S^{(0)}(u,{\bbeta})$ is incompletely observed and converges in probability to $s^{(0)}(u,\bbeta)$. We propose to replace $S^{(0)}(u,{\bbeta})$ with $\widehat{S}^{(0)}(u,\widehat{\bbeta},\widehat{\balpha})/\widehat{\lambda}_0(u)$, where $\widehat{\lambda}_0(\cdot)$ is the kernel smoothing estimator for the baseline function $\lambda_0(\cdot)$, defined as
$$\widehat{\lambda}_0(t) = \int_{0}^{\infty} \frac{\sum_{i=1}^nK_h(t-u) dO_i(u)}{\sum_{i=1}^{n}I(C_i\ge u)\exp\{\balpha^\top \bX_i(u) \}}. $$
It can be shown that $\widehat{S}^{(0)}(u,\widehat{\bbeta},\widehat{\balpha})/\widehat{\lambda}_0(u)$ converges in probability to $s^{(0)}(u,\bbeta_0)$. Thus we propose the following estimator for $\widehat{\mathcal{M}}_0(t)$,
\begin{align*}
\widehat{\mathcal{M}}_0(t) = \frac{1}{n}\int_{0}^t \frac{\sum_{i=1}^n\widehat{\lambda}_0(u) dN_i(u)}{ \widehat{S}^{(0)}(u,\widehat{\bbeta}, \widehat{\balpha}) }.
\end{align*} 
We show in the Supplementary Materials that, for each $t\in[0,\tau]$, $\sqrt{n}\{\widehat{\mathcal{M}}_0(t)-{\mathcal{M}}_0(t) \}$ converges in distribution to a zero mean normal random variable as $n\rightarrow\infty$.

\subsection{Estimation when VAR is violated}
The VAR assumption on the non-event visits may be violated in practice. In this section, we investigate the bias in estimating the proportional rate model under two special cases of VNAR. In the first scenario, the non-event visit time model and the event time model share the same set of time-dependent covariates. In the second scenario, the event and non-event visits depend on different sets of covariates. We show that the inclusion of informative non-event visits can bias the estimation of parameters of covariates shared by the event model and the non-event visit model, but may permit consistent estimation of other parameters.

In the first scenario, {the non-event visit process depends on the current value of $\bZ(t)$ and we assume}
\begin{align}
\label{m3}
E\{d{O}^*(t) \mid \bZ(t) \} = \exp\{  \bgamma_0^\top \bZ(t)  \}\lambda_0(t)dt, 
\end{align} 
where $\bgamma_0$ is a $p\times 1$ vector of regression coefficients and $\lambda_0$ is an unspecified baseline function. Since the coefficients $\bgamma_0$ cannot be directly estimated with the observed data, it is not clear how to apply the proposed method to estimate $\bE(t,\bbeta)$.  A naive approach is to use the unweighted estimator $\widehat{\bE}_O(t,\bbeta)$  in (\ref{k1}) and solve the estimating equation  $\bU_3(\bbeta) = n^{-1}\sum_{i=1}^{n} \int_{0}^{\tau} \{\bZ_i(t) - \widehat{\bE}_O(t,\bbeta)\} dN_i(t) = \bzero$ as if VCAR were true. For $k=0,1$, following the fact that
\begin{align*}
E[\bZ(t)^{k} \exp\{\bbeta^\top \bZ(t) \} d{O}(t)] 
= E\left[ I(C\ge t)\bZ(t)^{k} \exp\left\{(\bbeta+\bgamma_0)^\top \bZ(t) \right\} \right] \lambda_0(t)dt,
\end{align*}
$\bU_3(\bbeta)$ can be shown to converge  in probability to ${\bm \mU}_0(\bbeta+\bgamma_0)$. Moreover, since $\bbeta_0$ is the solution to ${\bm \mU}_0(\bbeta)=\bzero$, the estimator derived by solving $\bU_3(\bbeta)=\bzero$ converges in probability to $\bbeta_0-\bgamma_0$. Thus kernel smoothing with $\bZ(t)$ measured at informative non-event visit can lead to biased estimation of $\bbeta_0$ when $\bgamma_0\neq \bzero$. On the other hand, covariates that do not impact the non-event visit process will have values of $\bgamma_0$ being zero and their effects on the event visit process can therefore be consistently estimated.

In the second scenario, the chance of a non-event visit at time $t$ depends on another set of intermittently observed time-dependent covariates $\bW(t)$. Then it is possible to construct asymptotically unbiased estimators via the kernel smoothing approach. For ease of discussion, we assume that there is no overlap between $\bW(t)$ and $\bZ(t)$. Write $\widetilde{\bZ}(t) = (\bZ(t)^\top,\bW(t)^\top)^\top$, then
{the observed covariates are $\{\widetilde{\bZ}(t)dN(t), \widetilde{\bZ}(t)dO(t); t\ge 0 \}$.} The models for events and non-event visits are 
\begin{align}
\label{m4}
E\{d{N}^*(t) \mid \widetilde{\bZ}(t) \} = \exp\{  \bbeta_0^\top \bZ(t) \}\mu_0(t)dt,\\
\label{m5}
E\{d{O}^*(t) \mid \widetilde{\bZ}(t) \} = \exp\{  \btheta_0^\top \bW(t) \}\lambda_0(t)dt,
\end{align}
where $\btheta_0$ is a vector of regression coefficients. {Since $\bW(t)$  is not fully observed, the estimation procedure described in Section 3.1 cannot be applied.}
If $\btheta_0$ were known, along the same line as the estimation under VAR, one can solve the estimating equation $\bU_4(\bbeta,\btheta_0)  = \bzero$ to estimate $\bbeta_0$, where
\begin{align*}
\bU_4(\bbeta,\btheta) = \frac{1}{n}\sum_{i=1}^n \int_{0}^{\tau} \{\bZ_i(t) - \widehat{\bE}_1(t,\bbeta,\btheta) \} dN_i(t) = \bzero,
\end{align*}
and $$\widehat{\bE}_1(t,\bbeta,\btheta) = \frac{\sum_{i=1}^n \int_{0}^{\infty}K_h(t-s)\bZ_i(s)\exp\{ \bbeta^\top \bZ_i(s) -\btheta^\top \bW_i(s) \}dO_i(s)}{\sum_{i=1}^n \int_{0}^{\infty}K_h(t-s)\exp\{ \bbeta^\top \bZ_i(s) -\btheta^\top \bW_i(s) \}dO_i(s)}.$$
Similarly, if $\bbeta_0$ were known, we can exchange the two processes by treating $O^*(\cdot)$ as the event process and $N^*(\cdot)$ as the counting process for non-event visits. We then solve $\bU_5(\bbeta_0,\btheta)  = \bzero$ to estimate $\btheta_0$, where
\begin{align*}
\bU_5(\bbeta,\btheta)  = \frac{1}{n}\sum_{i=1}^n \int_{0}^{\tau} \{\bW_i(t) - \widehat{\bE}_2(t,\bbeta,\btheta) \} dO_i(t) = \bzero,
\end{align*}
and $$\widehat{\bE}_2(t,\bbeta,\btheta) = \frac{\sum_{i=1}^n \int_{0}^{\infty}K_h(t-s)\bW_i(s)\exp\{ \btheta^\top \bW_i(s)-\bbeta^\top \bZ_i(s)  \}dN_i(s)}{\sum_{i=1}^n \int_{0}^{\infty}K_h(t-s)\exp\{ \btheta^\top \bW_i(s)-\bbeta^\top \bZ_i(s)  \}dN_i(s)}.$$
Note that $\bU_4(\bbeta,\btheta)  = \bzero$ and $U_5(\bbeta,\btheta)  =\bzero$ provide a set of just-identified estimating equations for $(\bbeta_0,\btheta_0)$. Moreover, it can be shown that, for $k=4,5$, $\bU_k(\bbeta,\btheta)$ converges in probability to a limiting function ${\bm \mU}_k(\bbeta,\btheta)$ and ${\bm \mU}_k(\bbeta_0,\btheta_0)=\bzero$. Therefore, we can solve $\bU_4(\bbeta,\btheta)  = \bzero$ and $\bU_5(\bbeta,\btheta)  =\bzero$ to obtain a consistent estimate of $(\bbeta_0,\btheta_0)$. {We note that the above argument breaks down if $\bW(t)$ and $\bZ(t)$ share a common subset of covariates. In this situation, the effects of this common set of covariates are not identifiable, while the effects of other covariates remain identifiable. Also, when a covariate in $\bW(t)$ is highly correlated with another covariate in $\bZ(t)$, the estimated coefficients may have large variability.}

\section{Simulations}
We conducted a series of simulation studies to evaluate the finite-sample performance of the proposed estimator, the pseudo partial likelihood (PPL) estimator proposed by \cite{li2016recurrent}, and the last observation carried forward (LOCF) estimator under both VAR and VNAR scenarios. The covariates of the $i$th subject in the recurrent event model are $\bZ_i(t) = \{Z_{i1},Z_{i2}(t),Z_{i3}(t) \}$. The baseline covariate $Z_{i1}$ was generated from a uniform distribution on $[-0.5,0.5]$. The value of the time-dependent covariate process $Z_{i2}(t)$ was generated from a renewal process that alternates between states 0 and 1. We set $Z_{i2}(0) = 1$ with probability $0.5$. The duration of each state was generated from an exponential distribution with rate $\xi_i$, where $\xi_i$ follows a gamma distribution with mean 1 and variance 0.2. The covariate $Z_{i3}(t)= \sin(\pi t + w_{i1})$ is a continuous process with $w_{i1}$ being generated from the uniform distribution on $[0,2\pi]$. Due to the complex nature of EHR data, the visit processes may depend on variables that cannot be completely observed. Let $L_{i}(t) = \sin(\pi t + w_{i2})$  be a latent process that mimics time-varying characteristics that are not captured in the EHR (e.g., behavioral factors), where $w_{i2}$ was generated from the uniform distribution on $[0,2\pi]$. The event process $N_i^*(\cdot)$ was generated from a Poisson process with intensity function $\lambda_{10}(t) \exp\{ \beta_{\rm B}Z_{i1} + \beta_{\rm T1} Z_{i2}(t) + \beta_{\rm T2} Z_{i3}(t) + \gamma_1 L_{i}(t) - 1\}$. Integrating out the latent process, we obtain the proportional rate model $$\mu\{t\mid \bZ_i(t)\} = \mu_0(t) \exp\{\beta_{\rm B}Z_{i1} + \beta_{\rm T1} Z_{i2}(t) + \beta_{\rm T2} Z_{i3}(t) \},$$
where $\mu_0(t) = \lambda_{10}(t) E[\exp\{ \gamma_1 L_{i}(t) -1 \}]$ gives the baseline rate function. Set $V_{i0} = T_{i0} = 0$ and denote by $T_{ij}$ and $V_{ik}$ the time to the $j$th event visit and the time to the $k$th non-event visit, respectively. The non-event visit process $O_i^*(\cdot)$ is a Poisson process with intensity function
\begin{align}
\label{lambda_s}
\lambda_{20}(t)\exp\{\alpha_1 Z_{i1} + \alpha_2 X_{i2}(t) + \alpha_3 X_{i3}(t) + \alpha_4 Z_{i2}(t) + \alpha_5 Z_{i3}(t) + \gamma_2 L_{i}(t)  \},
\end{align}
where $X_{ij}(t) = Z_{ij}(\max(T_{i,N_i^*(t-)},V_{i,O_i^*(t-)}))$ is the last observed value of $Z_{ij}(t)$ before $t$ for $j = 2,3$. The non-event visit process is allowed to depend on the observed history, the current value $\bZ_i(t)$, and the latent process $L_{i}(t)$. When $(\alpha_4,\alpha_5) \neq (0,0)$, the VAR assumption is violated.

In what follows, we used $\bX_i(t) = \{Z_{i1}, X_{i2}(t), X_{i3}(t)\}$ as the covariates in (\ref{m2}), regardless of the true non-event visit model. We set $\beta_{B} = \beta_{T1} = -1$, $\beta_{T2} = 1$, $\lambda_{10}(t) = t$ and $\lambda_{20}(t) = 1$. Other parameters vary across scenarios and are specified in each scenario described below. The censoring time was generated from the uniform distribution on $[0,5]$. For each simulation, we generated 1000 simulated datasets, each with 200 subjects. The Epanechnikov kernel function $K(x) = 0.75(1-x^2)I(|x|<1)$ and bandwidth $h = 2n^{-1/3}$ were used. Results with a different bandwidth ($h = 2n^{-2/5}$) are given in the Supplementary Materials for evaluating the impact of different bandwidths. Nonparametric bootstrapping with individual as the sampling unit was applied to calculate the standard error of the proposed estimator.

\subsection{Simulations under VAR }

In the first set of simulations, we evaluated the performance of different methods when VAR is satisfied. We set $\alpha_4 = \alpha_5 = 0$ and consider scenarios corresponding to VCAR and VAR: \\
\indent (I) $\alpha_1 = \ldots = \alpha_3 = 0$, $\gamma_1 = \gamma_2 = 0$;\\
\indent (II) $\alpha_1 = \alpha_2 = -1/2$, $\alpha_3 = 1/2$, $\gamma_1 = \gamma_2 = 0$;\\
\indent (III) $\alpha_1 = \alpha_2 = -1$, $\alpha_3 = 1$, $\gamma_1 = \gamma_2 = 0$;\\
\indent (IV) $\alpha_1 = \ldots = \alpha_3 = 0$, $\gamma_1 = \gamma_2 = 1$;\\
\indent (V) $\alpha_1 = \alpha_2 = -1/2$, $\alpha_3 = 1/2$, $\gamma_1 = \gamma_2 = 1$;\\
\indent (VI) $\alpha_1 = \alpha_2 = -1$, $\alpha_3 = 1$, $\gamma_1 = \gamma_2 = 1$.\\
Under Scenarios I and IV, the VCAR assumption holds and the PPL approach should give consistent estimates of $\bbeta_0$. Under Scenarios II, III, V and VI, it can be shown that $E\{dO_i^*(t)\mid \bZ_i(t),\bX_i(t) \} = E\{dO_i^*(t)\mid \bX_i(t) \}$, thus the VAR assumption is satisfied. Under Scenarios I-IV, the Cox-type non-event visit model (\ref{m2}) holds and the proposed method provides consistent estimation of $\bbeta_0$. Under Scenarios V and VI, Model (\ref{m2}) is not the true model due to the correlation between $L_{i}(\cdot)$ and $\bX_i(\cdot)$, allowing us to investigate the performance of the proposed method when VAR is satisfied but the non-event visit model is misspecified.

Table 1 reports the summary statistics for the simulation studies. The proposed method performed well under Scenarios I-IV, and yielded almost negligible bias under Scenarios V and VI. Under {Scenarios} I and IV, the PPL approach performed well but had slightly larger variances than the proposed method. In other scenarios, the PPL approach yielded biased estimation, and the corresponding 95\% confidence intervals had poor coverage probability. {Compared to Scenario II (and V),  $|\alpha_j| ~(j=1,2,3)$ increases and the non-event visit process has a stronger degree of dependency on each covariate in Scenario III (and VI), resulting in larger biases in the PPL approach.} In all the scenarios, the LOCF approach yielded biased estimation and poor coverage probabilities for $\beta_{\rm T1}$ and $\beta_{\rm T2}$, while the estimation of $\beta_{\rm B}$ was almost unbiased and had smaller variance. It is worthwhile to point out that, when $Z_{i1}$ is correlated with $\{Z_{i2}(\cdot),Z_{i3}(\cdot)\}$, LOCF may yield biased estimation even for $\beta_{\rm B}$. With different bandwidths, the standard errors and the biases of the proposed estimator were similar (See Supplementary Materials). The results demonstrate that the performance of the proposed estimator is fairly stable with properly selected bandwidths.

{\linespread{1}
\begin{table}[thbp]
		\caption{Bias, asymptotic standard error estimates and confidence interval coverage probabilities for three estimators under VAR}
\label{tab1}
\begin{center}
	\begin{tabular}{crrrrrrrrrrrrrrr}
		\hline \hline \\[-2ex]
		    &   & \multicolumn{4}{c}{The proposed method} &\phantom{a} & \multicolumn{4}{c}{PPL}& &\multicolumn{4}{c}{LOCF}\\  
		\\[-2ex] 
		\cline{3-6}\cline{8-11} \cline{13-16}\\[-2ex]
		 Scenario & & Bias      & SE     & SEE     & CP         && Bias      & SE      &SEE    & CP    & \phantom{a}  & Bias      & SE     & SEE     & CP   \\
		\\[-2ex] 
		\hline
		\\[-2ex]
		I (VCAR) & $\beta_{\rm T1}$ & -3 & 21 & 22 & 97 && -3 & 21 & 22 & 97  && 50  & 13  & 13  & 4    \\
		& $\beta_{\rm T2}$  & 5  & 15 & 17 & 96 && 5  & 16 & 17 & 96 && -75 & 9   & 9   & 0   \\
		& $\beta_{\rm B}$& -4 & 33 & 35 & 97 && -5 & 38 & 39 & 96 && 2   & 22  & 22  & 95  \\

		II (VAR) & $\beta_{\rm T1}$  & -2 & 21 & 22 & 96 && 28  & 20 & 22 & 74  && 51  & 13  & 13  & 3  \\
		& $\beta_{\rm T2}$  & 3  & 16 & 17 & 96 && -13 & 15 & 16 & 84 && -75 & 9   & 9   & 0   \\
		& $\beta_{\rm B}$ & -2 & 33 & 35 & 96 && 57  & 34 & 37 & 64  && 1  & 23  & 22  & 92 \\
		
		III (VAR) & $\beta_{\rm T1}$ & -1 & 23 & 24 & 97 && 61  & 20 & 21 & 18  && 51  & 12  & 12  & 2    \\
		& $\beta_{\rm T2}$  & 5  & 17 & 17 & 96 && -31 & 14 & 15 & 44 && -73 & 10  & 9   & 0   \\
		& $\beta_{\rm B}$ & 8  & 39 & 38 & 94 && 117 & 33 & 35 & 8   && 2   & 21  & 21  & 95 \\
		
		IV (VCAR) & $\beta_{\rm T1}$  & -3 & 18 & 19 & 96 && -3 & 18 & 19 & 97 && 42  & 13  & 13  & 11 \\
		& $\beta_{\rm T2}$ & 4  & 13 & 14 & 97 && 4  & 13 & 14 & 96 && -59 & 10  & 9   & 0     \\
		& $\beta_{\rm B}$ & -5 & 29 & 31 & 96 && -4 & 33 & 34 & 95 && 3   & 22  & 21  & 93 \\
		
		V (VAR-M) & $\beta_{\rm T1}$ & -1 & 19 & 19 & 95 && 33  & 18 & 18 & 57  && 43  & 13  & 12  & 8  \\
		& $\beta_{\rm T2}$ & 3  & 13 & 14 & 96 && -21 & 13 & 13 & 65 && -59 & 10  & 9   & 0   \\
		& $\beta_{\rm B}$ & -1 & 27 & 30 & 97 && 56  & 29 & 32 & 56  && 3   & 22  & 21  & 95 \\

		VI (VAR-M) & $\beta_{\rm T1}$ & -3 & 20 & 20 & 95 && 66  & 17 & 18 & 5   && 42  & 12  & 12  & 8    \\
		& $\beta_{\rm T2}$ & 2  & 15 & 15 & 94 && -47 & 13 & 13 & 7 && -57 & 9   & 9   & 0    \\
		& $\beta_{\rm B}$ & 2  & 34 & 33 & 94 && 111 & 29 & 31 & 4   && 3   & 21  & 21  & 94 \\
\hline
	\end{tabular}
\end{center}
{\footnotesize
	Note: VAR-M stands for the case where VAR is satisfied but the non-event visit model is misspecified. PPL stands for the pseudo partial likelihood approach; LOCF stands for the last observation carried forward approach. Bias is the empirical bias ($\times 100$); SE is the empirical standard error ($\times 100$); SEE is the empirical mean of the standard error estimates; CP is the empirical coverage probability ($\times 100$) of the $95\%$ confidence interval.}	
\end{table}
}

\subsection{Simulations under VNAR}

In a second set of simulations, we conduct an additional set of simulation studies to evaluate the robustness of the proposed methods under VNAR scenarios. We set $\alpha_1 = -1$, $\gamma_1 =\gamma_2 = 1$ and consider the following scenarios:\\
\indent (VII) $\alpha_2 = -1$, $\alpha_4 = 0$, $\alpha_3 = \alpha_5 = 1/2$;\\
\indent (VIII) $\alpha_2 = \alpha_4 = -1/2$, $\alpha_3 = 1$, $\alpha_5 = 0$;\\
\indent (IX)  $\alpha_2 = \alpha_4 = -1/2$, $\alpha_3 = \alpha_5 = 1/2$;\\
\indent (X) $\alpha_2 = \alpha_3 = 0$, $\alpha_4 = -1/2$, $\alpha_5 = 1/2$.\\
Under Scenarios VII-X, the non-event visit process remains correlated with $\{Z_{i2}(t),Z_{i3}(t)\}$ after controlling for $\bX_i(t)$; thus VAR is violated. The simulation results are summarized in Table 2. All three methods yielded biased estimation, and the proposed method had the smallest bias in most scenarios. When the non-event visit process only depended on the current value of a single time-dependent covariate (Scenarios VII and VIII), the proposed estimation procedure yielded small biases and good coverage probabilities for the coefficients of the other covariates. Compared to PPL, the proposed method was more robust to model mis-specification. The LOCF approach had smaller variance but larger bias, resulting in poor coverage probabilities for both $\beta_{\rm T1}$ and $\beta_{\rm T2}$.

{\linespread{1}
\begin{table}[ht]
		\caption{Bias, asymptotic standard error estimates and confidence interval coverage probabilities for three estimators under VNAR}
\label{tab2}
\begin{center}
	\begin{tabular}{crrrrrrrrrrrrrrr}
		\hline \hline \\[-2ex]
		    &   & \multicolumn{4}{c}{The proposed method} &\phantom{a} & \multicolumn{4}{c}{PPL}& &\multicolumn{4}{c}{LOCF}\\  
		\\[-2ex] 
		\cline{3-6}\cline{8-11} \cline{13-16}\\[-2ex]
		 Scenario & & Bias      & SE     & SEE     & CP         && Bias      & SE      &SEE    & CP    & \phantom{a}  & Bias      & SE     & SEE     & CP   \\
		\\[-2ex] 
		\hline
		\\[-2ex]
		VII (VNAR)& $\beta_{\rm T1}$  & 32 & 20 & 21 & 64 && 85  & 18 & 19 & 1  && 45  & 12  & 12  & 6  \\
		& $\beta_{\rm T2}$ & 4  & 15 & 15 & 94 && -44 & 13 & 13 & 9  && -57 & 9   & 10  & 0   \\
		& $\beta_{\rm B}$ & -1 & 30 & 32 & 96 && 112 & 28 & 30 & 3  && 0   & 22  & 21  & 95 \\
		VIII (VNAR)& $\beta_{\rm T1}$ & -5  & 18 & 19 & 95 && 59  & 17 & 18 & 12 && 40  & 13  & 12  & 12   \\
		& $\beta_{\rm T2}$  & -40 & 14 & 14 & 22 && -73 & 13 & 13 & 0  && -60 & 10  & 10  & 0  \\
		& $\beta_{\rm B}$& -5  & 30 & 31 & 94 && 104 & 28 & 29 & 6  && -2  & 22  & 21  & 95  \\
	    IX (VNAR)& $\beta_{\rm T1}$   & 30  & 18 & 19 & 64 && 81  & 18 & 18 & 0  && 43  & 13  & 13  & 10\\
		& $\beta_{\rm T2}$  & -39 & 15 & 14 & 23 && -72 & 13 & 13 & 0  && -60 & 10  & 10  & 0  \\
		& $\beta_{\rm B}$ & -8  & 29 & 30 & 96 && 105 & 28 & 29 & 6  && -1  & 22  & 22  & 94 \\
		X (VNAR)& $\beta_{\rm T1}$ & 31  & 17 & 18 & 59 && 49  & 18 & 18 & 20 && 41  & 13  & 13  & 14   \\
		& $\beta_{\rm T2}$  & -39 & 13 & 13 & 16 && -49 & 13 & 13 & 6  && -58 & 10  & 10  & 0   \\
		& $\beta_{\rm B}$ & -4  & 24 & 28 & 97 && 100 & 28 & 30 & 8  && -3  & 23  & 22  & 93 \\\hline
	\end{tabular}
\end{center}
{\footnotesize
	Note: PPL stands for the pseudo partial likelihood approach; LOCF stands for the last observation carried forward approach. Bias is the empirical bias ($\times 100$); SE is the empirical standard error ($\times 100$); SEE is the empirical mean of the standard error estimates; CP is the empirical coverage probability ($\times 100$) of the $95\%$ confidence interval.}	
\end{table}
}

{In summary}, the performance of the proposed method depends on whether $\bZ(t)$ has additional effects on the non-event visits after controlling for $\bX(t)$. However, when $E\{d{O}^*(t) \mid \bX(t),\bZ(t)\}$ only depends on observed $\bX(t)$, the bias is relatively small if the Cox-type model can provide a reasonable approximation of $E\{d{O}^*(t) \mid \bX(t)\}$. When $E\{d{O}^*(t) \mid \bX(t),\bZ(t)\}$ depends on $\bZ(t)$, the proposed method leads to biased estimation. To achieve lower biases, we recommend researchers include all variables that may be predictive of the non-event visits in $\bX(t)$ and consider nonlinear spline terms to flexibly accommodate functional forms.

\section{Kidney Transplant Data Analysis}

The study cohort consists of 160 patients who underwent a kidney transplant at the Johns Hopkins Hospital in the year of 2012. Data were collected prospectively until September 24, 2014, that is, the date of data lock for our analysis. The time origin is set to be the date of the transplant. The median follow-up time was 20.2 months, with six deaths observed before September 24, 2014. Among the 160 study subjects, 41\% were females, 41\% had alive donors, 80\% were HLA compatible, and 53\% had donors with positive Cytomegalovirus (CMV) status. The ages ranged between 20 and 82 years, with an average of 52 years. By a thorough review of the relevant microbiology and clinical laboratory data, a total of 199 serious infection episodes were identified among the 654 clinical visits. Among them, 69 (35\%) were viral infections, and  130 (65\%) were infections involving gastrointestinal tract, bloodstream, and lower-respiratory tract. For kidney transplant recipients, serum creatinine concentration was routinely measured at clinical visits to monitor the kidney function. The average creatinine level at event visits was $2.46$ mg/dL, while the average creatinine level at non-event visits was $1.39$ mg/dL, lending credence to a less strong visit dependence for non-event visits. Figure 1 depicts the visit times as well as the creatinine levels measured at the event and non-event visits. It can be observed that the 124 patients who received either cadaveric or HLA incompatible donor organ were monitored more frequently and at a higher risk of serious infections (173 event visits and 386 non-event visits), compared to the 36 patients who received living and HLA compatible donor organ  (26 event visits and 67 non-event visits).

\begin{figure}[bh]
	\includegraphics[width = 1.1\textwidth]{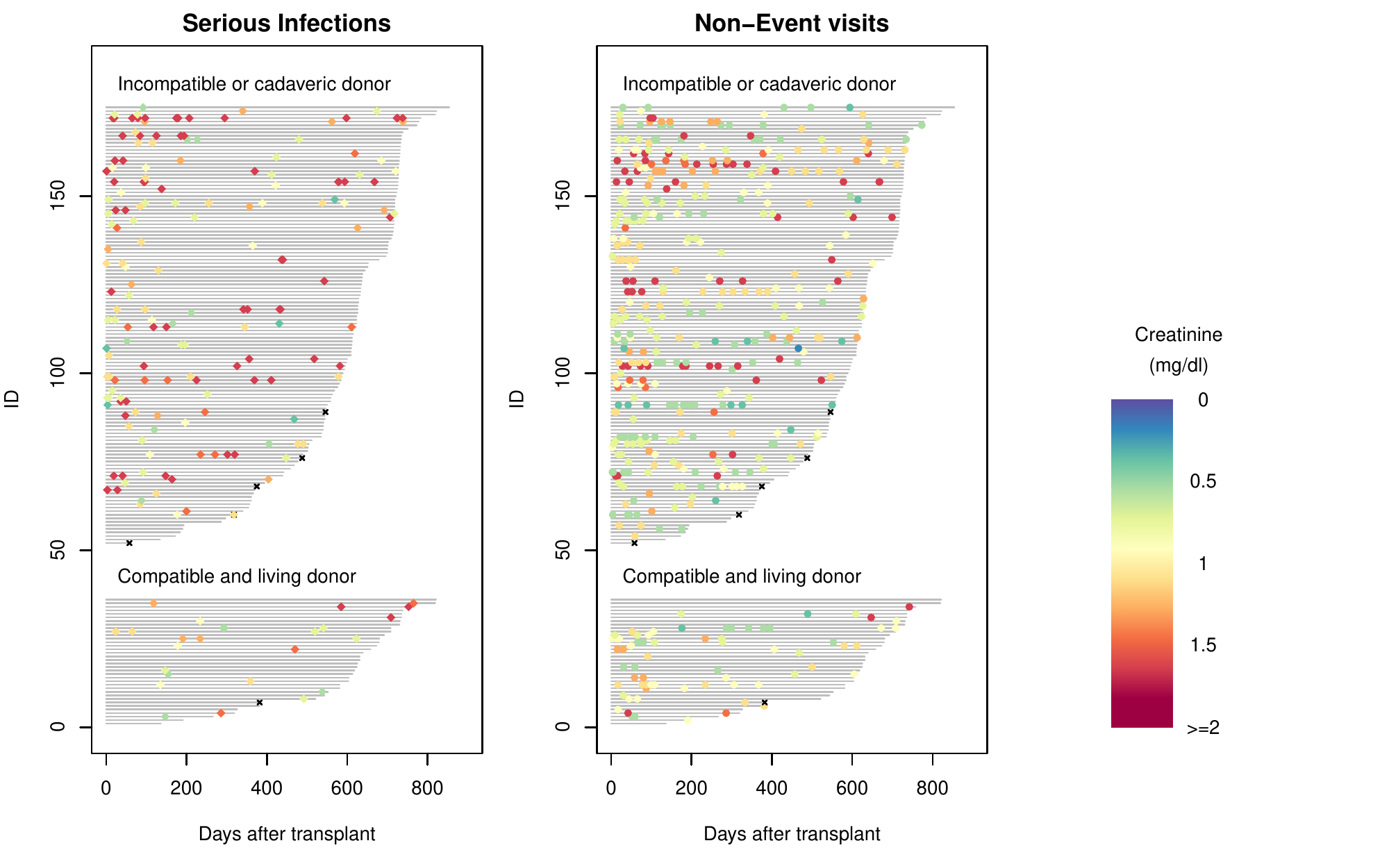}
	\caption{Longitudinal depiction of event and non-event visits}
	{\footnotesize
		Note: Each line segment corresponds to the duration of follow up of a patient, spanning from the day of transplant to the study end or loss to follow-up. The dots represent the visits after transplant and are colored by the creatinine level measured at the current visit. The black crosses indicate loss to follow-up due to death. }
\end{figure}

Let $N^*(t)$ represent the number of recurrent serious infections occurred at or before time $t$. To evaluate the effects of aforementioned demographic and clinical factors on the risk of experiencing serious infections, we impose the following semiparametric proportional rates model for the underlying recurrent event process  $N^*(t)$:
\begin{align*}
\mu\{t\mid \bZ(t) \} = \mu_0(t)\exp\{\beta_{\rm HLA}Z_1 + \beta_{\rm CAD}Z_2 + \beta_{\rm CMV}Z_3 +\beta_{\rm AGE}Z_4 + \beta_{\rm GEN}Z_5 +\beta_{\rm CREAT}Z_6(t) \},
\end{align*}
where $Z_1$ is a binary indicator such that $Z_1 = 1$ indicates HLA incompatible, $Z_2$ is a binary indicator such that $Z_2 = 1$ indicates cadaveric donor organ, $Z_3$ is a binary indicator such that $Z_3 = 1$ indicates donor is CMV positive, $Z_4$ is age in years (centered to have mean zero), $Z_5$ is a binary indicator such that $Z_5 = 1$ indicates male. Moreover, the binary covariate process $Z_6(t)=I(\text{serum creatinine level at time }t >2 \text{mg/dL})$ indicates an elevated creatinine level, where the cut off value was selected according to clinical recommendation \citep{morgan2007outcomes} for studying post-transplant outcomes. Note that the value of  $Z_6(t)$ can only be observed at post-transplant visits. {In addition to treating creatinine as a binary variable, we conducted another set of analysis by setting $Z_6(t)$ to be the log transformed creatinine level. The results are presented in the Supplementary Materials.}

\subsection{Estimating the non-event visit process}

Let $O^*(t)$ denote the number of non-event visits occurred before $t$. As can be observed in Figure 1, the non-event visit depends on the HLA compatibility status and the donor organ type of the transplantation. Moreover, because a high creatinine concentration is indicative of impaired renal function and thus might require closer monitoring of the patient, the scheduling of clinical visits also depends on whether the patient had an elevated serum creatinine level at the last (event or non-event) visit. 
For the non-event visits, we assume the non-event visits follows the following model,
\begin{align*}
E\{dO^*(t)\mid \bZ(t), \bX(t) \} = \lambda_0(t)\exp\left\{ \sum_{j=1}^5 \alpha_{0j}Z_j + \sum_{j=1}^ 6\alpha_j X_j(t)\right\}dt,
\end{align*}
where  $X_j(t) = I(O^*(t-)+N^*(t-)>0)Z_j$ for $j = 1,\ldots,5$, {$X_6(t) = I(O^*(t-)+N^*(t-)>0)\widetilde{Z}_6(t)$, and $\widetilde{Z}_6(t)$ denotes the value of creatinine concentration measured at the most recent visit prior to time $t$. }Note that the last measurement of creatinine level is available when a patient had a clinic visit before time $t$, that is, $O^*(t-)+N^*(t-)>0$.
Thus the imposed model allows the rate of the non-event visits to depend only on baseline factors if no clinical visit was recorded before time $t$, and to depend on both baseline factors and the last observed creatinine level if there is a visit before $t$. For $j = 1,\ldots,5$, the coefficient $\alpha_{0j}$ describes the effect of a baseline factor before any clinical visit, and the coefficient $\alpha_{0j}+\alpha_j$ characterizes its effect after the first clinical visits. The estimated regression coefficients are given in Table 3. {Compared to female patients, male patients had less frequent non-event visits ($62\% [\approx 1-\exp(-0.97)]$ and 52\% lower, respectively, before and after the first post-operative assessment). Older patients were monitored more frequently, and the effect of age slightly decreased after adjusting for the most recent assessment of creatinine. Donor status also played an important role in determining the non-event visits: after the first post-operative assessment, patients with HLA incompatible and cadaveric donors were monitored more frequently ($92.1\%$ and $62.7\%$ higher, respectively). Donor CMV status played a less important role, and its effect was not statistically significant. The effect of creatinine level from the most recent visit was relatively small and did not reach statistical significance.} {The estimated cumulative baseline function of the non-event process is shown in the left panel of Figure 2. It can be observed that the chance of a non-event visit was generally higher before day 100 and declined thereafter, indicating the patients were monitored more closely in the first three months. This is consistent with the current clinical practice for post-operative care of kidney transplant patients.}

{\linespread{1}
	\begin{table}[h]
		\small
		\caption{Estimated coefficients and 95\% confidence intervals for the non-event visit model of the kidney transplant study, before and after the first visit }
		\begin{center}
			\begin{tabular}{llclc}
				\hline \hline \\[-2ex]		
				& \multicolumn{2}{c}{Pre-first-visit} & \multicolumn{2}{c}{Post-first-visit}\\[1ex]
			HLA incompatible (yes vs. no) &	$\alpha_{01}$ & 0.39 & $\alpha_{01}+\alpha_1$ & 0.65 \\ 
			 & & (-0.18, 0.95) & & (0.22, 1.08)\\
			Cadaveric organ (yes vs. no)  &	$\alpha_{02}$ 	& 0.19 & $\alpha_{02}+\alpha_2$	& 0.49 \\
			& &  (-0.23, 0.61) & & (0.13, 0.85)\\
			Donor CMV (+ vs. --) &	$\alpha_{03}$ & 0.032 & $\alpha_{03}+\alpha_3$ &  -0.23 \\
			& &  (-0.33, 0.39)& & (-0.56, 0.10)\\
			Age &	$\alpha_{04}$ & 0.013 & $\alpha_{04}+\alpha_4$ & 0.009  \\
			& & (-0.001, 0.028)& & (-0.006, 0.024)\\
			Gender (male vs. female) &	$\alpha_{05}$ & -0.97  & $\alpha_{05}+\alpha_5$ & -0.73 \\
			& & (-1.35, -0.59)& & (-1.11, -0.35)\\
			Creatinine & & & $\alpha_{6}$ & -0.02 \\
			& & & &   (-0.56, 0.53)\\
				\hline
			\end{tabular}
		\end{center}
	\end{table}
}

\subsection{Estimating the event process}

To analyze the risk of serious infection, we consider three types of methods: the proposed method, the pseudo-partial-likelihood method \citep{li2016recurrent}, and the LOCF method. Because the baseline value of creatinine was not available, in the LOCF approach, we used the creatinine measured at the first visit to approximate the creatinine before the first visit. Table 4 summarizes the coefficient estimates and the 95\% confidence intervals. When the bandwidth $h$ was chosen as 60 days, patients with HLA incompatible donors and elevated creatinine level had a significantly higher risk of serious infection. Cadaveric donor, positive donor CMV status, elder age, and gender male were associated with higher infection risk, but the effects were not statistically significant. Moreover, using different bandwidths yielded similar conclusions. However, when applying PPL, cadaveric donor was associated with lower infection risk, the effects of HLA compatible transplant was not significant, and the effect of gender was significant. When applying LOCF, the effects of donor CMV status and age had different directions compared to the proposed method, although the effects were not statistically significant. {The estimated cumulative baseline function of the event process using the proposed method is shown in the right panel of Figure 2. For patients with the mean age 52, gender female, compatible HLA, alive donor, negative donor CMV status and relatively low creatinine levels over time, the expected number of serious infection within 700 days was slightly less than one. Moreover, the rate of serious infection decreased as time elapsed.}

{\linespread{1}
	\begin{table}[h]
		\small
		\caption{Estimated coefficients and 95\% confidence intervals for the event model of the kidney transplant study, fit by three methods and using different bandwidths, $h$ (days)}
		\begin{center}
			\begin{tabular}{lcccccc}
				\hline \hline \\[-2ex]
			&	  	\multicolumn{2}{c}{Proposed method}& \multicolumn{2}{c}{PPL} & \multicolumn{1}{c}{LOCF} \\
			&	  $h =  30$ & $h =  60$ & $h =  30$ & $h =  60$ & \\
				\hline                   
		$\beta_{\rm HLA}$	& 0.71 & 0.77 & 0.11  & 0.14 & 0.71\\
				& (-0.13, 1.56) &(0.006,1.55) &(-0.69, 0.92) &(-0.58, 0.86)&(0.18, 1.24)\\
		$\beta_{\rm CAD}$ 	& 0.21 & 0.26 & -0.24 & -0.19 & 0.37\\
			& (-0.44, 0.84)& (-0.32, 0.84) & (-0.93, 0.45)&(-0.81, 0.43) & (-0.012, 0.75)\\
		$\beta_{\rm CMV}$ 	& 0.065 & 0.095 & 0.21 & 0.25 & -0.058 \\
			 & (-0.48, 0.61) & (-0.39, 0.59) &(-0.41, 0.83) &(-0.32, 0.83) & (-0.42, 0.31)\\
		$\beta_{\rm AGE}$ 	&0.006 & 0.007 & -0.006& -0.005 & -0.001\\
			& (-0.015, 0.027)  & (-0.014, 0.027) &(-0.028, 0.016) &(-0.026, 0.016) & (-0.014,  0.011)\\
		$\beta_{\rm GEN}$ 	&0.21 & 0.21 & 1.02& 1.01  & 0.27 \\
			& (-0.35, 0.76) & (-0.29, 0.71) &(0.42, 1.62) &(0.47, 1.55) & (-0.11,  0.64)\\
		$\beta_{\rm CREAT}$ & 1.25 & 1.24 & 1.21 &1.21& 1.07 \\
			&(0.54, 1.96)	&(0.63, 1.86) & (0.41, 2.01)&(0.47, 1.94) & (0.69,  1.44) \\
		\hline
			\end{tabular}
		\end{center}
	\end{table}
}

\begin{figure}[h]
	\includegraphics[width = 0.5\textwidth]{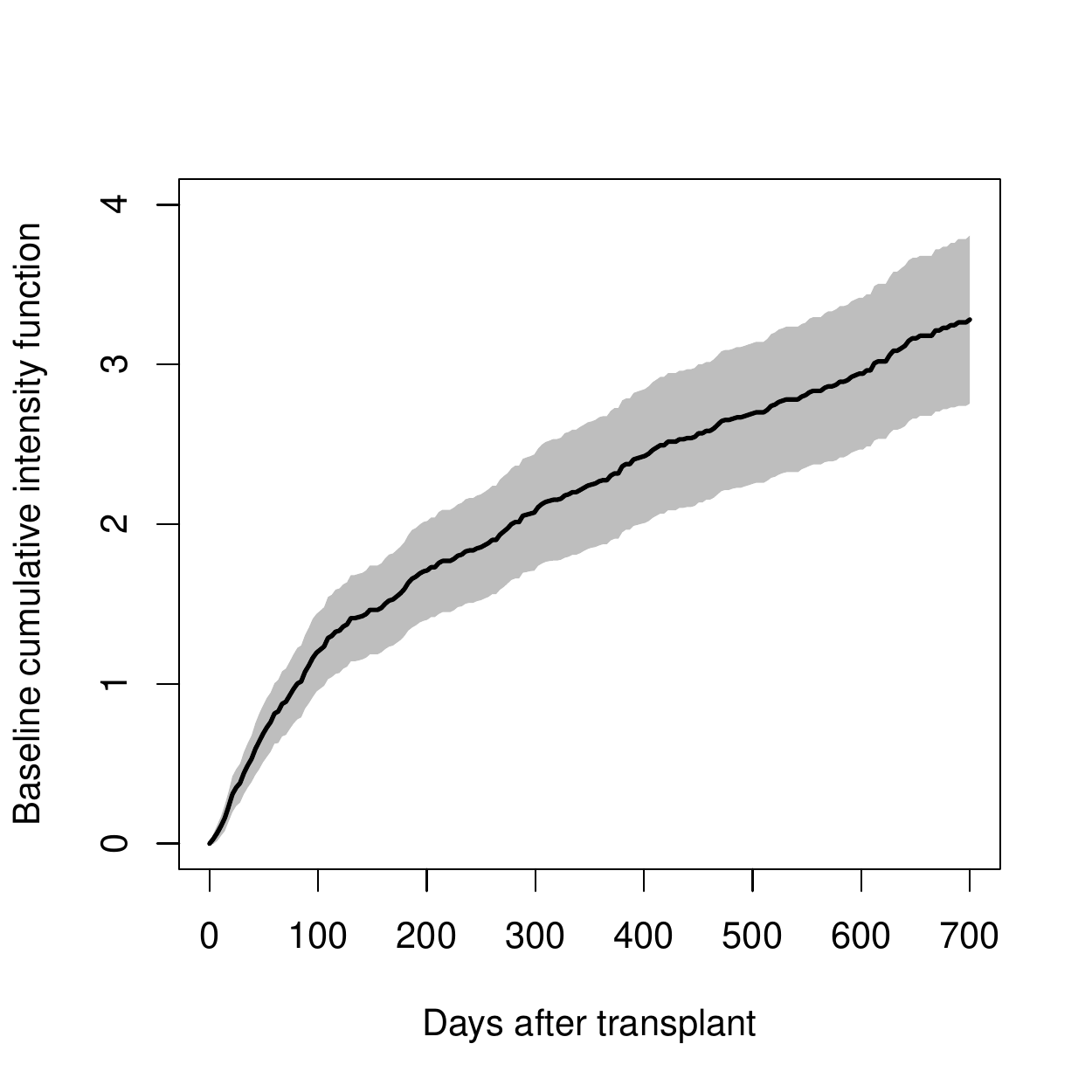}
	\includegraphics[width = 0.5\textwidth]{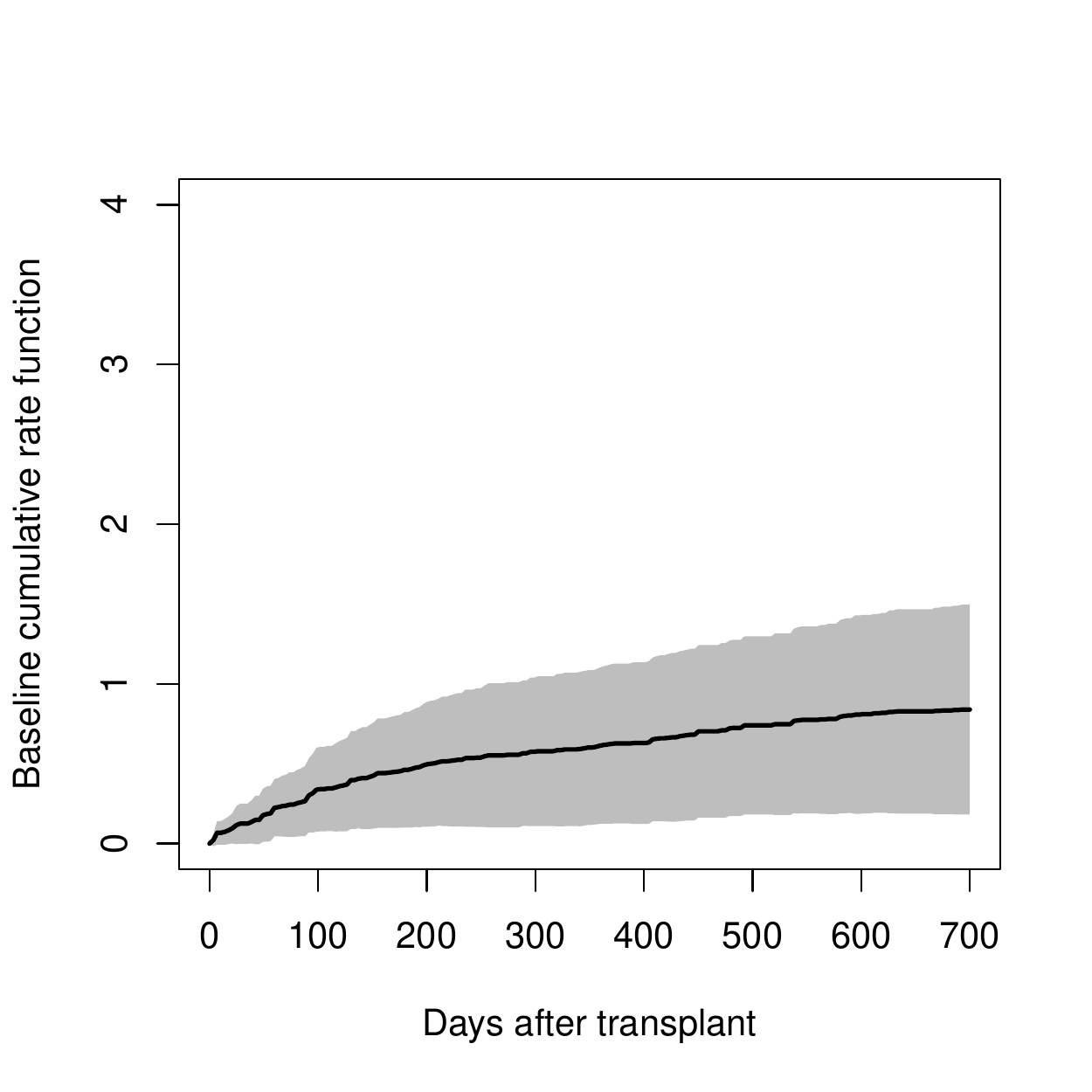}\\[-8ex]
	\caption{Baseline functions in non-event visit model (left) and event visit model (right)}
	{\footnotesize Note: The solid lines are the estimated baseline functions, and the shaded areas denote the pointwise 95\% confidence intervals.}
\end{figure}

\section{Discussion}

In this paper, we proposed a novel semiparametric approach to deal with intermittently measured covariates in the proportional rate model. Our method kernel smooths the mean functions of weighted covariate processes $\bE(t,\bbeta)$ and thus avoids modeling the covariate process. We require that covariate measurements from both non-event visits and event visits are available, but only covariates observed at non-event visits are used in kernel smoothing. 

{In biomedical studies, LOCF method is an easily-implemented and popular approach to deal with intermitently observed time-dependent covariates. When the time-dependent covariate $\bZ(t)$ does not vary much and can be densely observed, LOCF may perform reasonably well. In EHR data, $\bZ(t)$ is usually sparsely measured. As a result, the LOCF method is generally biased with poor confidence interval coverage probabilities, while the proposed method is essentially unbiased under the VAR assumption.}  

Our VAR assumption on the visit time process is suitable when the decision of a non-event visit is made based on the observed history (e.g., previous medical records). {In addition to applications in medical science, the VAR framework is potentially useful in other applications where the observation of time-dependent covariates are initiated by study investigators based on observed subject characteristics. For example, the method can be applied to study risk factors for recurrent child maltreatments \citep{hindley2006risk}. Here, the non-event visits can be home visits, and home visitors conduct an assessment of the family and record the possibly time-dependent risk factors at each visit. The frequency of home visits usually depends on the previous history of maltreatments, thus the VAR assumption is reasonable. As another example, the method may also be applied to study risk factors for recurrent food-borne illness of restaurants, where the non-event visits are health inspections initiated by the Department of Health based on, for example, types of food preparation and history of food safety violations.} In practice, if the decision to visit at $t$ depends on current value of $\bZ(t)$ conditional on $\bX(t)$, the proposed method may yield biased estimation. Methods that deal with VNAR will be investigated in our future research work.

In this article, we considered the case where all the time-dependent covariates are measured at each event and non-event visit. It is worthwhile to point out that in the recurrent event analysis setting, the measurements of time-dependent covariates are usually available at all the event visits. If the covariates are only observed at non-event visits, we can extend the proposed methods by adding another additional layer of kernel smoothing along the same line as \cite{cao2015analysis}. {Finally, the current discussion focused on time-dependent covariates such as biomarkers or treatments. In practice, the EHR data contain enormous information such as measurements as images and curves \citep{de2016functional}, and clinical notes that describe patients' condition \citep{wu2016omic}. Inclusion of more complex time-dependent covariates will be investigated in our future work.}

\section*{Acknowledgement}

This research was partially supported by NIH R01CA193888. The transplant study was supported in part by NIH K24AI085118. The first author's research was partially supported by the Calderone Junior Faculty Prize from Columbia University Mailman School of Public Health.

\begin{appendix}
	\begin{center}
		{\normalsize \textbf{APPENDIX}}
	\end{center}
	\medskip
We assume the following regularity conditions for Theorem 3.1.
\begin{itemize}
	\item[(A1)] $\{\mN^*_i(C_i),\mO^*_i(C_i),{\bm \mZ}^*_{{\rm obs},i}(C_i),C_i,i=1,\ldots,n \}$ are independent and identically distributed.
	\item[(A2)] The true parameter $\bbeta_0$ lies in a compact set $\mathcal{B}$ in $\mathbb{R}^p$, and the true parameter $\balpha_0$ lies in a compact set ${\mathcal{A}}$ in $\mathbb{R}^q$.
	\item[(A3)] $N(\tau)$ is bounded. The rate function of $N(t)$ is of bounded variation for $t\in[0,\tau]$.
	\item[(A4)] The covariate processes $\bZ(t)$ and $\bX(t)$ are left continuous and have right-hand limits, and they have bounded total variation. 
	\item[(A5)] For $k = 0,1$, the function $\bs^{(k)}(t,\bbeta,\balpha)= E[I(C\ge t)\bZ(t)^k\exp\{\bbeta^\top \bZ(t)-(\balpha-\balpha_0)^\top \bX(t) \}]$ has bounded second order derivative for $t\in[0,\tau'],\bbeta\in{\mathcal B}, \balpha\in{\mathcal A}$, where $\tau'$ is a constant such that $\tau'>\tau$. Moreover, $s^{(0)}(t,\bbeta)>0$ for $t\in[0,\tau']$.
	\item[(A6)] The process $O(t)$ is bounded for $t\in[0,\tau']$. The baseline function $\lambda_0(\cdot)$ is positive and has bounded second order derivative for $t\in[0,\tau']$.
	\item[(A7)] The kernel function $K(\cdot)$ is a symmetric density function on $[-1,1]$.
	\item[(A8)] $h\propto n^{-\nu}$, where $1/4<\nu<1/2$.
\end{itemize}

	\section*{Proof of Theorem 3.1}
	\setcounter{equation}{0}
	\renewcommand{\thesection}{A}
	\setcounter{subsection}{0}
	\setcounter{lemma}{0}
	\renewcommand{\thelemma}{A\arabic{lemma}}
We first prove the consistency of $\widehat{\bbeta}$. Applying the results of \cite{lin2000semiparametric}, $\widehat{\balpha}$ converges in probability to $\balpha_0$. Because the functional defined by $\bU_2$ in (\ref{ee2}) is continuous with respect to the supremum norm topology, it is sufficient to show that the four processes $n^{-1}\sum_{i=1}^n\int_{0}^{\tau} \bZ_i(t)dN_i(t)$, $n^{-1}\sum_{i=1}^nN_i(t)$, $\widehat{\bS}^{(1)}(t,\bbeta,\widehat{\balpha})$, $\widehat{S}^{(0)}(t,\bbeta,\widehat{\balpha})$ converge in probability to their limits uniformly for $\beta\in\mathcal{B}$ and $t\in[0,\tau]$. By the Law of large numbers, $n^{-1}\sum_{i=1}^n\int_{0}^{\tau} \bZ_i(t)dN_i(t)$ converges in probability to $E\int_{0}^{\tau} \bZ(t)dN(t)$. Let $\mu_c(t)$ be the marginal rate function of $N(t)$, that is, $E\{dN(t) \} = \mu_c(t)dt$. Because $n^{-1}\sum_{i=1}^n N_i(t)$ is a bounded monotone process, it converges in probability to $\int_{0}^{t}\mu_c(u)du$. In what follows, we first show the uniform consistency of $\widehat{\bS}^{(k)}(t,\bbeta,\widehat{\balpha})$ for $k = 0,1$.

For $k = 0,1$ and $t\in[h,\tau]$, define $\bB_i^{(k)}(t,\bbeta,\balpha) = \int_{0}^{t}\bZ_i(u)^k\exp\{ \bbeta^\top \bZ_i(u)-\balpha^\top \bX_i(u) \} dO_i(u)$, then $\widehat{\bS}^{(k)}(t,\bbeta_0,\balpha_0) = \sum_{i=1}^n\int_0^\infty K_h(t-u)\bB_i^{(k)}(du,\bbeta_0,\balpha_0)/n$. Moreover, we set $\bB_i^{(k)}(t,\bbeta_0,\balpha_0) = \bB_i^{(k)}(h,\bbeta_0,\balpha_0)$ for $t\in[0,h)$. Then $\widehat{\bS}^{(k)}(t,\bbeta,\balpha) = \int_{0}^{\infty} K_h(t-u) \widehat{E}\{ \bB^{(k)}(du,\bbeta,\balpha)\}$. For $k = 0,1$, the function class $\{ \bB^{(k)}(t,\bbeta,\balpha), t\in[0,\tau], \bbeta\in\mathcal{B}, \balpha\in\mathcal{A} \}$ has $L_2$ bracketing number of polynomial order. By straightforward algebra, we have 
\begin{align*}
\sup_{t\in[0,\tau],\bbeta\in\mathcal{B},\balpha\in\mathcal{A}}|\widehat{\bS}^{(k)}(t,\bbeta,\balpha)-E\{ \widehat{\bS}^{(k)}(t,\bbeta,\balpha)\}|  \le  h^{-1} |\widehat{E}\{ \bB^{(k)}(t,\bbeta,\balpha)\}-{E}\{ \bB^{(k)}(t,\bbeta,\balpha)\} |V(K).
\end{align*}
where $V(K)$ is the variation of kernel function $K$. 
By Theorem 2.14.9 in \cite{vandervaart}, for some constants $c_1,c_2,c_3>0$, we have
\begin{align*}
P\left(\sup_{t\in[0,\tau],\bbeta\in\mathcal{B},\balpha\in\mathcal{A}} \sqrt{n}|\widehat{E}\{ \bB^{(k)}(t,\bbeta,\balpha)\}-{E}\{ \bB^{(k)}(t,\bbeta,\balpha)\} | > x\right)< c_1 x^{c_2} e^{-c_3x^2}.
\end{align*}
When $nh^2\rightarrow\infty$, we have
\begin{align*}
P\left(\sup_{t\in[0,\tau],\bbeta\in\mathcal{B},\balpha\in\mathcal{A}} h^{-1}|\widehat{E}\{ \bB^{(k)}(t,\bbeta,\balpha)\}-{E}\{ \bB^{(k)}(t,\bbeta,\balpha)\} | > \epsilon \right)< c_1 (\sqrt{n}h\epsilon)^{c_2} e^{-c_3nh^2} \rightarrow 0.
\end{align*}
Thus $\sup_{t\in[0,\tau],\bbeta\in\mathcal{B},\balpha\in\mathcal{A}}|\widehat{\bS}^{(k)}(t,\bbeta,\balpha)-E\{ \widehat{\bS}^{(k)}(t,\bbeta,\balpha)\}| =o_p(1)$. 

Define $R(t) = I(C\ge t)$ and $$\bs^{(k)}(t,\bbeta,\balpha) = E[R(t)\bZ(t)^k\exp\{\bbeta^\top \bZ(t)-(\balpha-\balpha_0)^\top \bX(t) \}].$$ 
Since the function $\bs^{(k)}(\cdot,\bbeta,\balpha) \lambda_0(\cdot)$ has bounded second order derivative on $[0,\tau']$, there exists a constant $M$ such that $|\bs^{(k)}(u,\bbeta,\balpha) \lambda_0(u) - \bs^{(k)}(t,\bbeta,\balpha) \lambda_0(t) - {\bm s}^{(k,1)}(t,\bbeta,\balpha) (u-t)|  \le M(u-t)^2$, where ${\bm s}^{(k,1)}(t,\bbeta,\balpha)={{\rm d}\{\bs^{(k)}(t,\bbeta,\balpha)\lambda_0(t)\}}/{{\rm d}t}$. For $t\ge h$, 
\begin{eqnarray*}
 \lefteqn{\Big|E\{ \widehat{\bS}^{(k)}(t,\bbeta,\balpha)\} -\bs^{(k)}(t,\bbeta,\balpha) \lambda_0(t)\Big|}\\
&= & \Big|\int_0^\infty K_h(t-u) \bs^{(k)}(u,\bbeta,\balpha) \lambda_0(u)du- {\bm s}^{(k,1)}(t,\bbeta,\balpha)\int_0^\infty K_h(t-u) (u-t)du-\bs^{(k)}(t,\bbeta,\balpha) \lambda_0(t)\Big|\\
&\le & M\int_0^\infty K_h(t-u) (u-t)^2 du = Mh^2 \kappa_2(K),
\end{eqnarray*}
where $\kappa_2(K) = \int x^2 K(x)dx$. Thus we have $\sup_{t\in[h,\tau],\bbeta\in\mathcal{B},\alpha\in\mathcal{A}}|E\{ \widehat{\bS}^{(k)}(t,\bbeta,\balpha)-\bs^{(k)}(t,\bbeta,\balpha)\lambda_0(t)\}| =O(h^2)$. Moreover, it can be shown that $\sup_{t\in[0,h],\bbeta\in\mathcal{B},\balpha\in\mathcal{A}}|\bs^{(k)}(h,\bbeta,\balpha)\lambda_0(h) -\bs^{(k)}(t,\bbeta,\balpha)\lambda_0(t)\}| =O(h)$ and $$\sup_{t\in[0,\tau],\bbeta\in\mathcal{B},\balpha\in\mathcal{A}}|\widehat{\bS}^{(k)}(t,\bbeta,{\balpha})-\bs^{(k)}(t,\bbeta,{\balpha})\lambda_0(t) | = o_p(1).$$
Therefore, 
\begin{eqnarray*}
\lefteqn{\sup_{t\in[0,\tau],\bbeta\in\mathcal{B}}|\widehat{\bS}^{(k)}(t,\bbeta,\widehat{\balpha})-\bs^{(k)}(t,\bbeta,\balpha_0)\lambda_0(t) |} \\
&\le & \sup_{t\in[0,\tau],\bbeta\in\mathcal{B}}|\widehat{\bS}^{(k)}(t,\bbeta,\widehat{\balpha})-\bs^{(k)}(t,\bbeta,\widehat{\balpha})\lambda_0(t) | + \sup_{t\in[0,\tau],\bbeta\in\mathcal{B}}|\bs^{(k)}(t,\bbeta,\widehat{\balpha})\lambda_0(t)-\bs^{(k)}(t,\bbeta,\balpha_0)\lambda_0(t) | \\
&\le & \sup_{t\in[0,\tau],\bbeta\in\mathcal{B},\balpha\in\mathcal{A}}|\widehat{\bS}^{(k)}(t,\bbeta,{\balpha})-\bs^{(k)}(t,\bbeta,{\balpha})\lambda_0(t) | + \sup_{t\in[0,\tau],\bbeta\in\mathcal{B}}|\bs^{(k)}(t,\bbeta,\widehat{\balpha})\lambda_0(t)-\bs^{(k)}(t,\bbeta,\balpha_0)\lambda_0(t) |\\
&= & o_p(1).
\end{eqnarray*}
The last inequality is due to the Continuous Mapping Theorem. Applying Lemma 2.1 in \cite{nan2013general}, we have $\widehat{\bbeta}\overset{p}{\rightarrow}\bbeta_0$.

We next prove the asymptotic normality of $\widehat{\bbeta}$. Define $\bet^{(k)}(t,\balpha_0) = E[R(t)\bX(t)^{\bigotimes k}\exp\{\balpha_0^\top \bX(t) \}]$, where $\ba^{\bigotimes 0} = 1$, $\ba^{\bigotimes 1} = \ba$, and $\ba^{\bigotimes 2} = \ba\ba^\top$. The large-sample property of the estimating equation $\bU_1(\balpha)=\bzero$ has been studied in \cite{lin2000semiparametric}. Following \cite{lin2000semiparametric}, we have $\sqrt{n}(\widehat{\balpha}-\balpha_0 ) = n^{-1/2}\sum_{i=1}^n {\bm \psi}_i + o_p(1)$, where 
	$${\bm \psi}_i = \bD_1^{-1} \int_{0}^{\tau} \left\{ \bX_i(t) -\frac{\bet^{(1)}(t,\balpha_0)}{\bet^{(0)}(t,\balpha_0)} \right\}dM_i(t,\balpha_0),$$ 
	$$\bD_1 = \int_{0}^{\tau} \left[ \frac{\bet^{(2)}(t,\balpha_0)}{\bet^{(0)}(t,\balpha_0)}-\left\{\frac{\bet^{(1)}(t,\balpha_0)}{\bet^{(0)}(t,\balpha_0)}\right\}^{\bigotimes 2} \right]dE{O}(t),$$ 
and $dM_i(t,\balpha_0) = d{O}_i(t) -  R_i(t)\exp\{\balpha_0^\top \bX(t) \}\lambda_0(t)dt$. \\
For the second set of estimating equations $\bU_2(\bbeta) = 0$, we have
\begin{eqnarray*}
	\lefteqn{ n^{-1/2}\bU_2(\bbeta_0) }\\
& = & n^{-1/2}\sumi \int_{0}^{\tau} \{\bZ_i(u) - \widehat{\mathcal{\bE}}(u,\bbeta_0,\widehat{\balpha})\} dN_i(u)\\
& = & n^{-1/2}\sumi \int_{0}^{\tau} \left\{\bZ_i(u) - \widehat{\mathcal{\bE}}(u,\bbeta_0,{\balpha_0})\right\} dN_i(u) + n^{-1/2}\sumi \int_{0}^{\tau} \left\{ \widehat{\mathcal{\bE}}(u,\bbeta_0,{\balpha_0})-\widehat{\mathcal{\bE}}(u,\bbeta_0,\widehat{\balpha}) \right\} dN_i(u) \\
&\overset{def}{=} & I + II.
	\end{eqnarray*}
Define $E\{ dN(t)\} = \mu_c(t)dt$. We first derive the i.i.d. representation of $I$.
\begin{align*}
I &= n^{-1/2}\sumi \int_{0}^{\tau} \left\{\bZ_i(u) - {\mathcal{\bE}}(u,\bbeta_0)\right\} dN_i(u) + n^{-1/2}\sumi \int_{0}^{\tau} \left\{{\mathcal{\bE}}(u,\bbeta_0) - \widehat{\mathcal{\bE}}(u,\bbeta_0,{\balpha_0})\right\} dN_i(u).
\end{align*}
Moreover, we have
\begin{eqnarray*}
\lefteqn{ n^{-1/2}\sumi \int_{0}^{\tau} \left\{ \widehat{\mathcal{\bE}}(u,\bbeta_0,{\balpha_0})-{\mathcal{\bE}}(u,\bbeta_0) \right\} dN_i(u) }\\
& = & n^{-1/2}\sumi \int_{0}^{\tau} \left\{ \widehat{\mathcal{\bE}}(u,\bbeta_0,{\balpha_0})-{\mathcal{\bE}}(u,\bbeta_0) \right\} \mu_c(u)du + o_p(1)\\
& = & n^{-1/2}\sumi \int_{0}^{\tau} \left\{ \widehat{\mathcal{\bE}}(u,\bbeta_0,{\balpha_0})-\frac{\bs^{(1)}(u,\bbeta_0)\lambda_0(u)}{\widehat{S}^{(0)}(u,\bbeta_0,\balpha_0)} \right\} \mu_c(u)du + \\
& & n^{-1/2}\sumi \int_{0}^{\tau} \left\{ \frac{\bs^{(1)}(u,\bbeta_0)\lambda_0(u)}{\widehat{S}^{(0)}(u,\bbeta_0,\balpha_0)}-{\mathcal{\bE}}(u,\bbeta_0) \right\} \mu_c(u)du + o_p(1)\\
&\overset{def}{=}& I_A + I_B + o_p(1).
\end{eqnarray*}
Along the same arguments as (i) and (ii), page 3060, in \cite{li2016recurrent}, let $g(t)$ be a nonnegative function of bounded variation on $[0,\tau]$, then for $k=0,1$, when $h \propto n^{-\nu}$ and $1/4<\nu<1/2$, we have
\begin{align}
\label{eq1}
\sqrt{n}\int_{0}^{\tau} g(t)\widehat{\bS}^{(k)}(t,
\bbeta_0, \balpha_0) dt - n^{-1/2}\sumi \int_{0}^\tau g(t) R_i(t)\bZ_i^k(t)\exp\{\bbeta_0^\top \bZ_i(t)-\balpha_0^\top \bX_i(t) \}dO_i(t) = o_p(1),
\end{align} 
and 
\begin{align}
\label{eq2}
\sqrt{n}\int_{0}^\tau g(t)\left\{\widehat{S}^{(0)}(t,\bbeta_0,\balpha_0)^{-1} - s^{(0)}(t,\bbeta_0)^{-1}\lambda_0(t)^{-1}  \right\}\left\{\widehat{\bS}^{(k)}(t,\bbeta_0,\balpha_0) - \bs^{(k)}(t,\bbeta_0)\lambda_0(t) \right\}dt = o_p(1).
\end{align}
The results can be obtained by replacing the observation process $dO(t)$ in \cite{li2016joint} with the weighted observation process $\exp\{-\balpha_0^\top \bX(t) \}dO(t)$.
Define $\bE(t,\bbeta) = \bs^{(1)}(t,\bbeta)/s^{(0)}(t,\bbeta)$. Based on the above results, we have
\begin{align*}
I_A =& n^{-1/2}\sumi \int_{0}^{\tau} \left\{ \widehat{\mathcal{\bE}}(u,\bbeta_0,{\balpha_0})-\frac{\bs^{(1)}(u,\bbeta_0)\lambda_0(u)}{\widehat{S}^{(0)}(u,\bbeta_0,\balpha_0)} \right\} \mu_c(u)du\\
= &  n^{-1/2}\sumi \int_{0}^{\tau} \left\{ \frac{\widehat{\bS}^{(1)}(u,\bbeta_0,\balpha_0)-\bs^{(1)}(u,\bbeta_0)\lambda_0(u)}{s^{(0)}(u,\bbeta_0)\lambda_0(u)} \right\} \mu_c(u)du + o_p(1)\\
= & n^{-1/2}\sumi \int_{0}^{\tau} \frac{\mu_c(t)}{\bs^{(0)}(t,\bbeta_0)\lambda_0(t)}\bB_i^{(1)}(dt, \bbeta_0, \balpha_0) - n^{1/2}\int_{0}^{\tau} \frac{\bs^{(1)}(t,\bbeta_0)\mu_c(t)}{s^{(0)}(t,\bbeta_0)}dt + o_p(1).
\end{align*}

\begin{align*}
I_B = & -n^{-1/2}\sumi \int_{0}^{\tau} \bs^{(1)}(u,\bbeta_0)\lambda_0(u)\left\{ \frac{\widehat{\bS}^{(0)}(u,\bbeta_0,\balpha_0)-s^{(0)}(u,\bbeta_0)\lambda_0(u)}{s^{(0)}(u,\bbeta_0)^2\lambda_0(u)^2}\right\}\mu_c(u)du + o_p(1)\\
= & -n^{-1/2}\sumi \int_{0}^{\tau} \bs^{(1)}(u,\bbeta_0)\lambda_0(u)\left\{ \frac{\widehat{S}^{(0)}(u,\bbeta_0,\balpha_0)}{s^{(0)}(u,\bbeta_0)^2\lambda_0(u)^2}\right\}\mu_c(u)du + n^{1/2}\int_{0}^{\tau} \frac{\bs^{(1)}(t,\bbeta_0)\mu_c(t)}{s^{(0)}(t,\bbeta_0)}dt+o_p(1)\\
= & -n^{-1/2}\sumi \int_{0}^{\tau}\frac{\bE(t,\bbeta_0)\mu_c(t)}{\lambda_0(t)}B_i^{(0)}(dt,\bbeta_0,\balpha_0) + n^{1/2}\int_{0}^{\tau} \frac{\bs^{(1)}(t,\bbeta_0)\mu_c(t)}{s^{(0)}(t,\bbeta_0)}dt+o_p(1).
\end{align*}
Therefore, we have
	\begin{align*}
	I = & n^{-1/2}\sumi \left[\int_{0}^{\tau} \{\bZ_i(t) - {\mathcal{\bE}}(t,\bbeta_0)\} dN_i(t) - \right.\\
	&\left. \int_{0}^{\tau} \frac{\bZ_i(t)-\mathcal{\bE}(t,\bbeta_0 )}{s^{(0)}(t,\bbeta_0)\lambda_0(t)} \exp\{\bbeta_0^\top \bZ_i(t)-\balpha_0^\top \bX_i(t)\}\mu_c(t)dO_i(t) \right] + o_p(1).
	\end{align*}	
We next derive the i.i.d. representation of $II$. Note that the $p\times q$ matrix $\frac{\partial \widehat{\mathcal{\bE}}(t,\bbeta_0,\balpha)}{\partial\balpha}$ is 
\begin{align*}
\frac{\partial \widehat{\mathcal{\bE}}(t,\bbeta_0,\balpha)}{\partial\balpha} =& -\frac{\int_{0}^\infty K_h(t-u) \widehat{E}[\bZ(u)\bX(u)^\top\exp\{\bbeta_0^\top \bZ(u)-\balpha^\top \bX(u) \}dO(u)] \widehat{S}^{(0)}(t,\bbeta_0,\balpha) }{\widehat{S}^{(0)}(t,\bbeta_0,\balpha)^2}+\\
& \frac{\widehat{\bS}^{(1)}(t,\beta_0,\alpha)\int_{0}^\infty K_h(t-u) \widehat{E}[\bX(u)^\top\exp\{\bbeta_0^\top \bZ(u)-\balpha^\top \bX(u) \}dO(u)] }{\widehat{S}^{(0)}(t,\bbeta_0,\balpha)^2}. 
\end{align*}
Moreover, it can be shown that
\begin{align*}
\frac{\partial \widehat{\mathcal{\bE}}(t,\bbeta_0,\balpha)}{\partial\balpha}\Big|_{\balpha = \balpha_0} \overset{p}{\rightarrow}-\frac{E[R(t)\bZ(t)\bX(t)^\top\exp\{\bbeta_0^\top \bZ(t) \} ]}{s^{(0)}(t,\bbeta_0)} + \bE(t,\bbeta_0)E[R(t)\bX(t)^\top\exp\{\bbeta_0^\top \bZ(t) \} ].
\end{align*}
Define $\mathcal{\bE}(t,\bbeta,\balpha) = \frac{E[ R(t)\bZ(t)^{k}\exp\{\bbeta^\top \bZ(t)-(\balpha-\balpha_0)^\top \bX(t) \}] }{E[R(t) \exp\{\bbeta^\top \bZ(t)-(\balpha-\balpha_0)^\top \bX(t) \}]}$, then $\mathcal{\bE}(t,\bbeta,\balpha_0) = \mathcal{\bE}(t,\bbeta)$. We have
	\begin{align*}
	II = &n^{1/2}\int_{0}^{\tau} \left\{ \widehat{\mathcal{\bE}}(t,\bbeta_0,{\balpha_0})-\widehat{\mathcal{\bE}}(t,\bbeta_0,\widehat{\balpha}) \right\} \mu_c(t)dt + o_p(1)\\
	= & - \int_{0}^\tau \frac{\partial \widehat{\mathcal{\bE}}(t,\bbeta_0,\balpha)}{\partial\balpha}\Big|_{\balpha = \balpha_0} \mu_c(t)dt\cdot n^{1/2}(\widehat{\balpha}-\balpha_0) + o_p(1)\\
	= & n^{-1/2}\sumi  \int_{0}^{\tau}\frac{E[R(t)\bZ(t)\bX(t)^\top\exp\{ \bbeta_0^\top \bZ(t) \}]}{s^{(0)}(t,\bbeta_0)}\mu_c(t)dt\cdot \bpsi_i-\\
	&n^{-1/2}\sumi \int_{0}^{\tau}\bE(t,\bbeta_0 )E[R(t)\bX(t)^\top\exp\{ \bbeta_0^\top \bZ(t) \}]\mu_c(t)dt \cdot \bpsi_i + o_p(1).
	\end{align*}
	Therefore, we have proved $\sqrt{n}\{\bU_2(\widehat{\bbeta})-\bU_2(\bbeta_0)\} = n^{-1/2}\sumi \bphi_i + o_p(1)$, where
	\begin{align*}
	\bphi & = \int_{0}^{\tau} \left\{ \bZ(t) - {\bE}(t,\bbeta_0)\right\}dN(t) - \int_{0}^{\tau} \frac{\bZ(t)-{\bE}(t,\bbeta_0 )}{s^{(0)}(t,\bbeta_0)\lambda_0(t)} \exp\{\bbeta_0^\top \bZ(t)-\balpha_0^\top \bX(t)\}\mu_c(t)dO(t) +\\
	& \int_{0}^{\tau} \frac{E[R(t)\bZ(t)\bX(t)^\top\exp\{ \bbeta_0^\top \bZ(t) \}]-\bs^{(1)}(t,\bbeta_0 )E[R(t)\bX(t)^\top\exp\{ \bbeta_0^\top \bZ(t) \}]}{s^{(0)}(t,\bbeta_0 )} \mu_c(t)dt\cdot \bpsi. 
	\end{align*}
	Following similar arguments as in \cite{li2016recurrent}, $\sqrt{n}(\widehat{\bbeta}-\bbeta_0)$ converges in distribution to $N(\bzero,\bD^{-1} {\bm V} (\bD^{\top})^{-1})$, where
	\begin{align*}
	\bD &= \int_{0}^{\tau} \left[ \left\{\frac{\bs^{(1)}(t,\bbeta_0)}{s^{(0)}(t,\bbeta_0)} \right\}^{\bigotimes 2}-\frac{\bs^{(2)}(t,\bbeta_0)}{s^{(0)}(t,\bbeta_0)}\right] \mu_c(t)dt,\\
	{\bm V} &= E( \bphi^{\bigotimes 2} ).
	\end{align*}
\end{appendix}

\bibliographystyle{asa}	
\bibliography{unblinded}
\end{document}